\documentclass{natureprintstyle}

\usepackage[para,symbol]{footmisc}
\usepackage{graphicx}
\usepackage{aas_macros} 
\usepackage{amssymb}
\usepackage{amsmath}
\usepackage{color}
\usepackage{url}
\usepackage{soul}
\usepackage{hyphenat}
\usepackage{hyperref}
\hypersetup{
    colorlinks,
    citecolor=blue,
    filecolor=blue,
    linkcolor=blue,
    urlcolor=blue
}
\setstcolor{red}

\usepackage[displaymath, mathlines]{lineno}
\renewcommand{\figurename}{{\bf Figure}}

\newcommand{\Ms} {\ifmmode{~{\rm M_\odot}}\else{M$_\odot$}\fi}
\newcommand{\Msyr} {\ifmmode{~{\rm M_\odot~yr^{-1}}}\else{M$_\odot$~yr$^{-1}$}\fi}
\renewcommand{\apj} {Astrophys. J.}
\renewcommand{\apjl} {Astrophys. J. Lett.}
\renewcommand{\apjs} {Astrophys. J. Suppl.}
\renewcommand{\mnras} {Mon. Not. R. Astron. Soc.}
\renewcommand{\nat} {Nature}

\renewcommand{\aap} {Astron. Astrophys.}
\newcommand{\nar} {New Astronomy Reviews}
\newcommand{\na} {New Astron.}

\def\change#1{{\color{black} #1}}
\def\moved#1{{\color{black} #1}}

\title{\nohyphens{Formation of massive black holes in rapidly growing
  pre-galactic gas clouds}}

\author{John ~H. Wise$^{1}$,
  John A. Regan$^2$,  Brian W. O'Shea$^{3,4}$,
  Michael L. Norman$^{5,6}$, Turlough P. Downes$^2$ \&
  Hao Xu$^{5,6,7}$}
\begin{document}

\maketitle
\begin{affiliations}
\item Center for Relativistic Astrophysics, School of Physics, Georgia
  Institute of Technology, Atlanta, GA 30332, USA;
  \texttt{jwise@gatech.edu}
\item Centre for Astrophysics and Relativity, School of Mathematical
  Sciences, Dublin City University, Dublin, Ireland
\item Department of Computational Mathematics, Science and
  Engineering, Michigan State University, East Lansing, MI 48824, USA
\item Department of Physics and Astronomy, Michigan State University,
  East Lansing, MI 48824, USA
\item Center for Astrophysics and Space Sciences, University of
  California, San Diego, La Jolla, CA 92093, USA
\item San Diego Supercomputer Center, San Diego, La Jolla, CA 92093,
  USA
\item IBM, 2455 South Road, Poughkeepsie, NY 12601
\end{affiliations}

\begin{abstract}

  The origin of supermassive black holes (SMBHs) that inhabit the
  centers of massive galaxies is largely
  unconstrained\cite{Volonteri_2012, Greif_2015}.  Remnants from
  supermassive stars (SMSs) with masses around 10,000 solar masses
  provide the ideal seed candidates, known as direct collapse black
  holes\cite{Omukai_2001, Begelman_2006, Hosokawa_2012, Ardaneh_2018}.
  However, their very existence and formation environment in the early
  Universe are still under debate, with their supposed rarity further
  exacerbating the problem of modeling their ab-initio
  formation\cite{Habouzit_2016, Chon_2018}.  SMS models have shown
  that rapid collapse, with an infall rate above a critical value, in
  metal-free haloes is a requirement for the formation of a
  proto-stellar core which will then form an SMS\cite{Hosokawa_2013,
    Umeda_2016}.  Using a radiation hydrodynamics simulation of early
  galaxy formation\cite{OShea_2015, Xu_2016}, we show the natural
  emergence of metal-free haloes both massive enough, and with
  sufficiently high infall rates, to form an SMS.  We find that haloes
  that are exposed to both a Lyman-Werner intensity of $J_{\rm LW}
  \sim 3\ J_{21}$\footnote{$J_{21}$ is the intensity of background
    radiation in units of $10^{-21} \rm ~erg ~cm^{-2} ~s^{-1} ~Hz^{-1}
    ~sr^{-1}$.} and that undergo at least one period of rapid growth
  early in their evolution are ideal cradles for SMS formation. This
  rapid growth induces substantial dynamical
  heating\cite{Yoshida_2003a, Fernandez_2014}, amplifying the existing
  Lyman-Werner suppression originating from a group of young galaxies
  20 kiloparsecs away. Our results strongly indicate that structure
  formation dynamics, rather than a critical Lyman-Werner (LW) flux,
  may be the main driver of massive black hole formation in the early
  Universe.  We find that massive black hole seeds may be much more
  common in overdense regions of the early Universe than previously
  considered with a comoving number density up to $10^{-3}
  \rm{Mpc}^{-3}$.

\end{abstract}


\change{Standard cold dark matter cosmologies predict large-scale
  structure forms hierarchically.  Smaller objects forming at early
  times subsequently merge and grow into larger objects.  The
  existence of SMBHs\cite{Mortlock_2011, Banados_2018} with masses
  around $10^9\Ms$ (\Ms, solar mass) only 800~Myr after the Big Bang
  indicate that there must have been an early intense convergence of
  mass in rare locations.}


We performed a suite of cosmological radiation hydrodynamics
simulations (named Renaissance; see Methods) to elucidate the
formation of the first generations of stars and galaxies in the
Universe\cite{OShea_2015, Xu_2016} with the code {\sc
  Enzo}\cite{Enzo_2014}.  It includes models for the formation of
massive metal-free (Population III; Pop III) stars and subsequent
metal-enriched stars not unlike ones found in the Galaxy.  We follow
the impact of their ionizing radiation\cite{WiseAbel_2011} and
supernova explosions on their environments as galaxies first assemble,
both of which play an important role in regulating early galaxy
formation.

Motivated by possible early SMS formation, we analyze the region from
the Renaissance Simulation suite that is centered on the densest
cosmological volume of 133.6 comoving Mpc$^3$ and contains 822
galaxies at its ending redshift of $z = 15$ (270 Myr after the Big
Bang).  We identify candidate SMS host haloes by searching the
simulation for metal-free atomic cooling haloes without prior star
formation at $z = 15$.  We place no constraints on the level of LW
flux impacting the haloes.  There are 670 atomic cooling haloes, ten
of which are metal-free and have not hosted prior star formation (see
Extended Data Table \ref{tab:haloes}).  The remaining atomic cooling
haloes have formed stars prompted by either H$_2$ or metal line
cooling and are not conducive for SMS (and subsequent direct collapse
black hole) formation.  Out of these ten haloes, we concentrate on two
``target'' haloes---the most massive halo (MMH) and, separately,
the most irradiated halo (LWH)---in this study.  We resample their
mass distributions at $z=20$ at a mass resolution higher by a factor
of 169 and resimulate them to study their gravitational collapse in
more detail.


\begin{figure*}[t]
  \includegraphics[width=\textwidth]{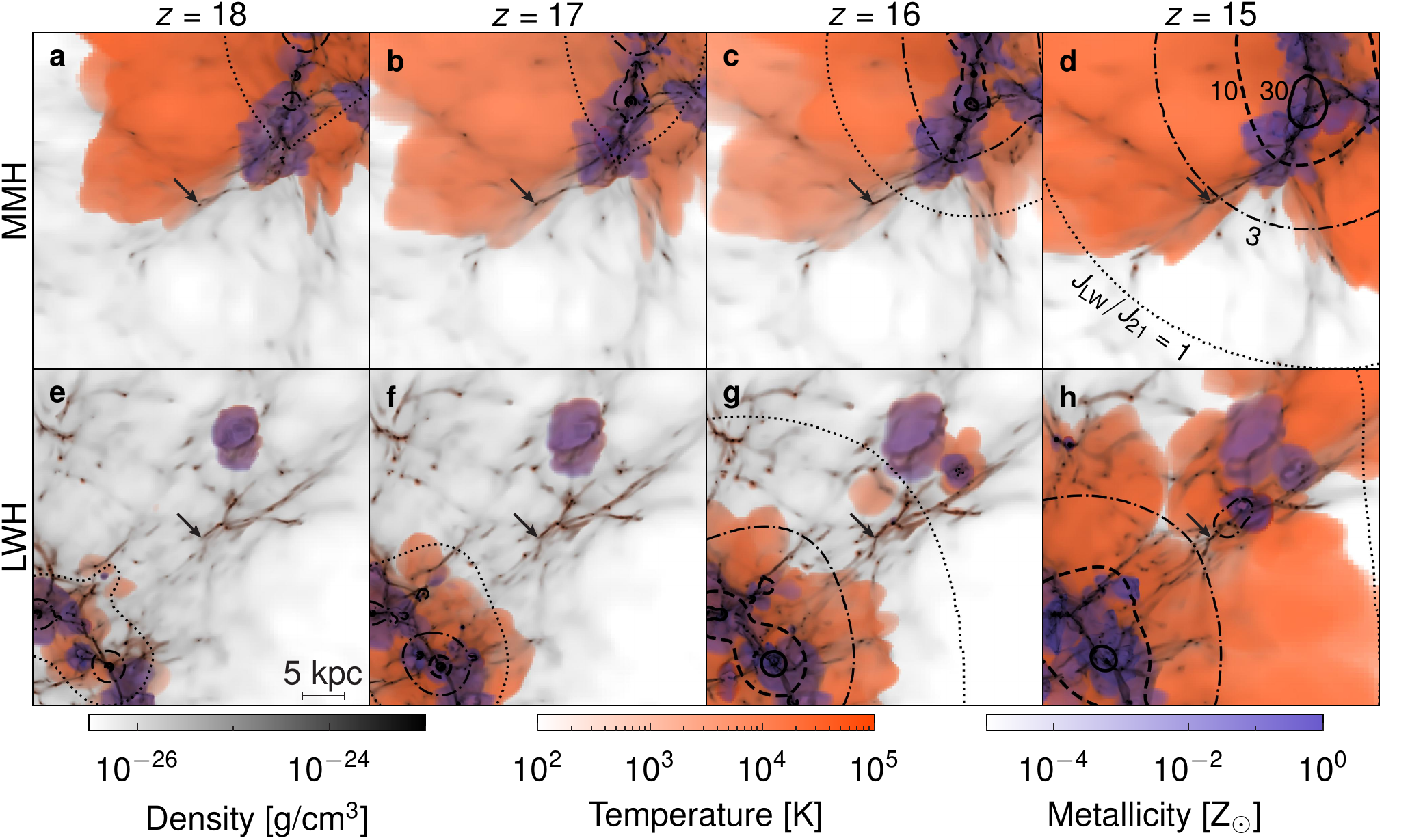}
  \caption{\textbf{Thermal and chemical evolution of the immediate
      pre-galactic environment.} Projections of temperature (orange),
    metallicity (blue), and gas density (black) of a region (40 kpc
    across with a 8 kpc depth) centered on the MMH ({\bf a-d}) and
    LWH ({\bf e-h}) that are highlighted with arrows.  Going from
    left to right shows the heated and metal-enriched volumes around
    early galaxies and Pop III stars growing from redshift 18
    to 15 (62 million years).  The dotted, dash-dotted, dashed, and
    solid contours indicate where the average Lyman-Werner flux is 1,
    3, 10, and 30 $J_{21}$.  Both candidate haloes that host massive
    black hole formation have $J_{21} \simeq 3$, are just outside of
    the cosmological H~{\sc ii} region, and are still unaffected by
    any external metal-rich winds.}
  \label{fig:composite}
\end{figure*}

\begin{figure}[t]
  \centering
  \includegraphics[width=\columnwidth]{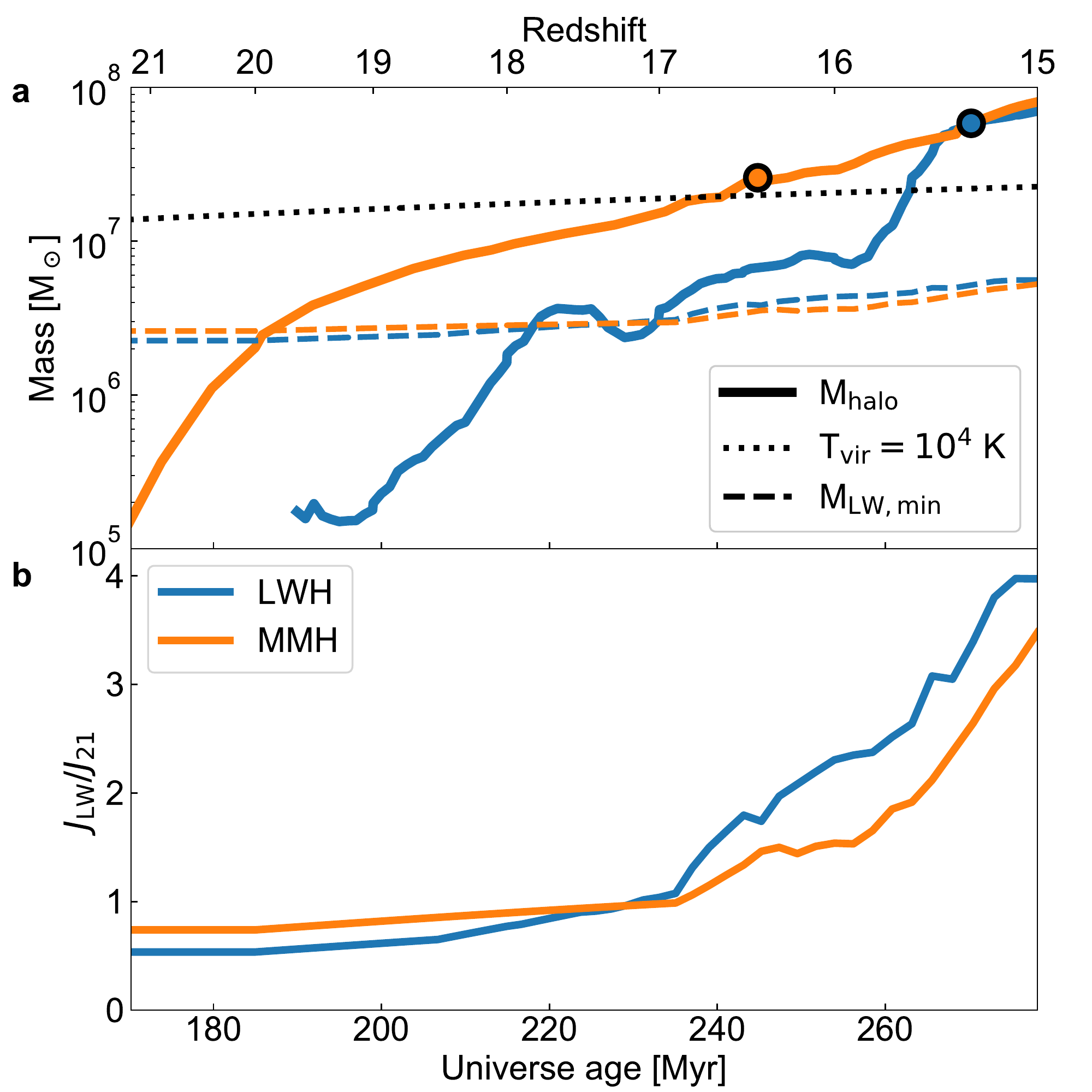}
  \caption{\textbf{Target halo mass growth histories.} The evolution
    of the metal-free atomic cooling halo with the highest mass
    (orange) and highest Lyman-Werner flux (blue).  {\bf a,} Halo mass
    (solid) compared with the minimum mass for H$_2$ cooling (dashed)
    and atomic cooling (dotted).  The circles indicate when the haloes
    experience gravitational collapse into a massive black hole.  The
    MMH grows smoothly and has a growth history more
    typical of an atomic cooling halo.  In contrast, the LWH
    undergoes rapid growth just before collapse,
    growing by a factor of six within 10 million years. {\bf b},
    Lyman-Werner flux at the densest point in units of $J_{21}$ that
    is increasing from the nearby growing group of young galaxies a
    distance of 10-15 kpc.}
  \label{fig:evo}
\end{figure}

\begin{figure*}[t]
  \centering
  \includegraphics[width=\textwidth]{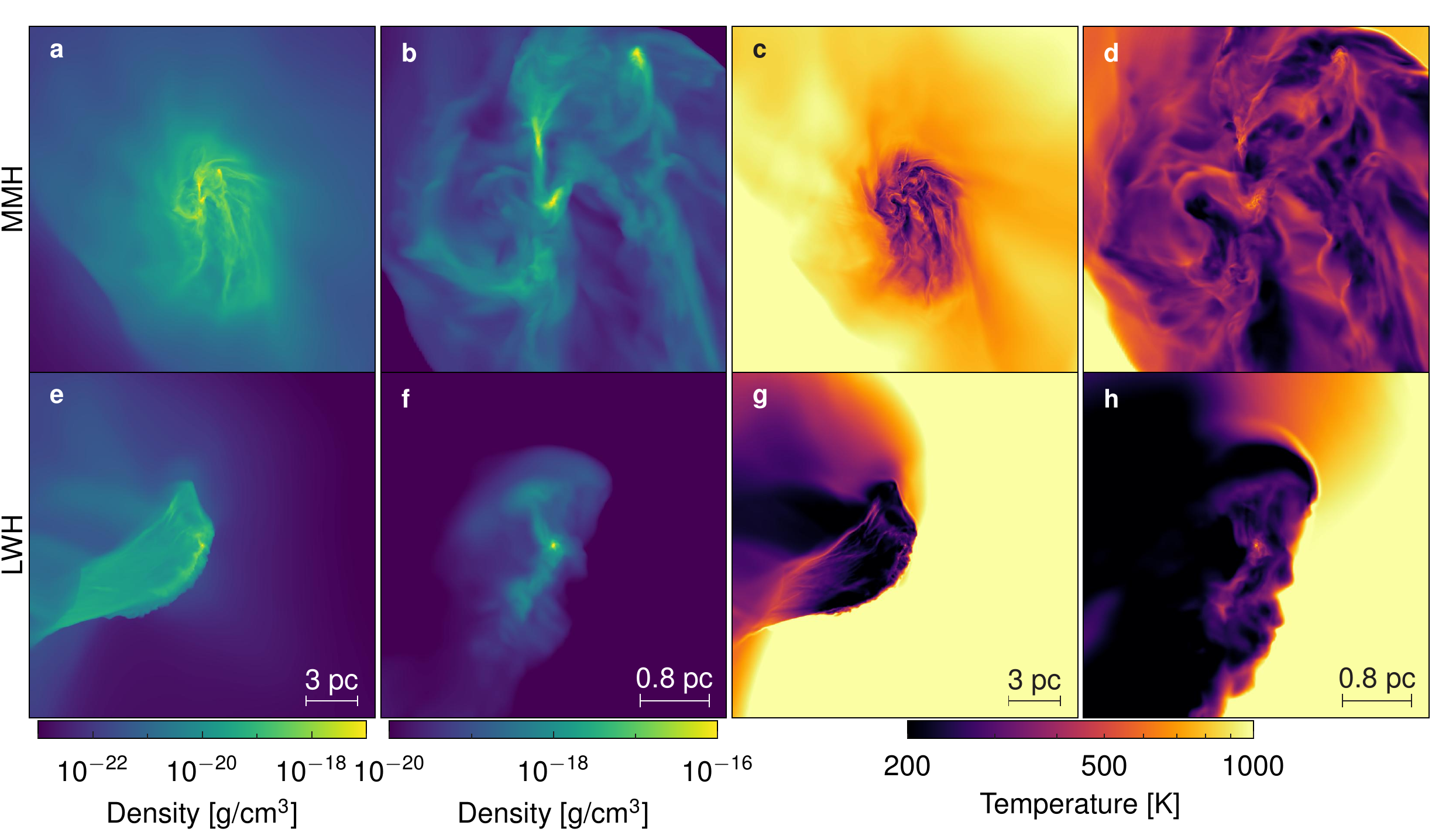}
  \caption{\textbf{Morphology of the collapsing objects.}
    Density-weighted projections of gas density (\textbf{a,b,e,f}) and
    temperature (\textbf{c,d,g,h}) of the MMH (top row) and LWH
    (bottom row) at the end of the simulations, centered on the
    densest point and aligned to be perpendicular with the angular
    momentum vector of the disk.  The MMH forms a cold disk that
    fragments into three clumps, accompanied by weak turbulent shocks
    within the disk.  In the LWH, a dense and cold sheet forms
    after the collision of two progenitor haloes, where a single clump
    collapses after becoming self-gravitating.  Within a radius of 0.1
    parsecs in all of the clumps, adiabatic compression heats the
    gas.}
  \label{fig:proj}
\end{figure*}

\begin{figure*}[t]
  \centering
  \includegraphics[width=\textwidth]{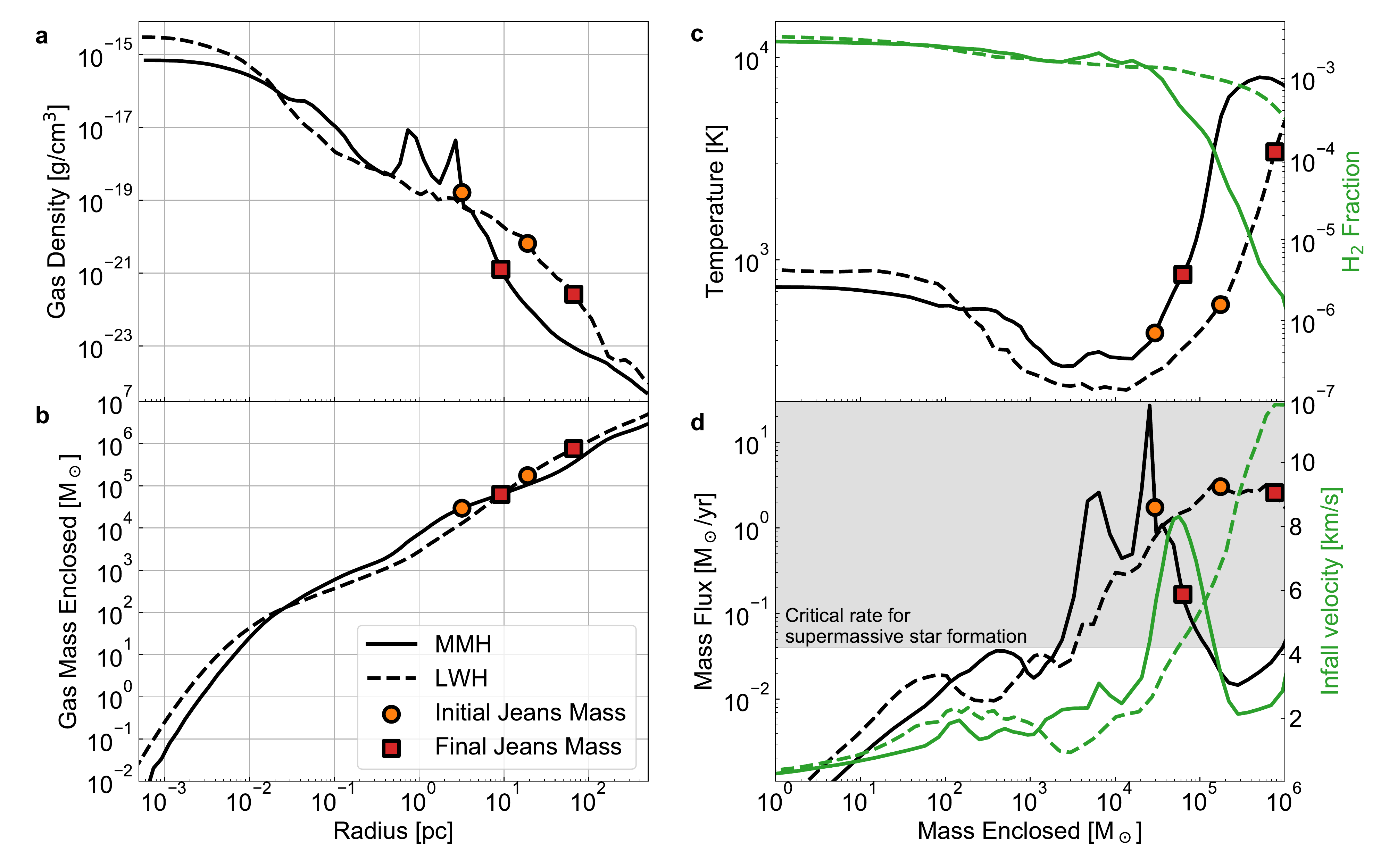}
  \caption{\textbf{Gravitational collapse of the target haloes.}
    Radially averaged profiles of gas density ({\bf a}), gas mass
    enclosed ({\bf b}), gas temperature ({\bf c}, black lines), H$_2$
    fraction ({\bf c}, green lines), radial mass infall rates ({\bf
      d}, black lines), and radial infall velocities ({\bf d}, green
    lines) in the MMH (solid) and LWH (dashed).  The orange
    circles indicate the Jeans mass and associated quantities at that
    mass scale when the object first becomes gravitationally unstable,
    whereas the red squares represent the same quantities at the end
    of the simulation.  At the surface of the gravitationally unstable
    core, gas is infalling at rates upwards of 0.1 solar masses per
    year, where \change{values above 0.04 solar masses per year can
      induce and fuel supermassive star formation.}}
  \label{fig:radial}
\end{figure*}

Both haloes assemble in a region 10--25~kpc away from a group of young
galaxies that have photo-ionized, photo-heated, and chemically
enriched their adjacent environments (Fig. \ref{fig:composite}).  At
$z = 18$, the young galaxies near the MMH
(Fig. \ref{fig:composite}a through \ref{fig:composite}d) have created
an amorphous H~{\sc ii} region with a maximum extent of 20~kpc.  As the
star formation rates grow in these young (massive) galaxies, the LW
intensities increase from $J_{\rm LW} \simeq 1~J_{21}$ at $z=18$
within 5~physical kpc of the galaxies to $30~J_{21}$ at $z=15$.  The
only other source of LW radiation comes from four nearby haloes
hosting Pop III stars, 3--5~kpc from the LWH.  Both target haloes
are impacted by a LW intensity $J_{\rm LW} \simeq 3~J_{21}$ at $z=15$
(Fig. \ref{fig:composite}d, \ref{fig:composite}h). This total flux
impinging on the target haloes is 6--600 times lower than previous
critical estimates for SMS formation\cite{Shang_2010, Agarwal_2015b,
  Glover_2015a}.

The metal-enriched regions around these galaxy congregations only
reach a distance of 5~kpc, far from the target haloes.  These heavy
elements originate from both their Pop III progenitors and ongoing
star formation in the galaxies.  Over the next 60 Myr, the
ionizing radiation from the young growing galaxies near both target
haloes extends the H~{\sc ii} regions to nearly 40 kpc in radius, evident
in Fig. \ref{fig:composite}.  This leaves the vast majority of the
intergalactic medium and associated collapsed haloes chemically
pristine but nonetheless bathed in LW radiation, helping to prevent
Pop III star formation.

During the halo assembly process (Fig. \ref{fig:evo}a), the LW
intensity increases from $0.3~J_{21}$ to $3~J_{21}$ at $z=15$
(Fig. \ref{fig:evo}b), corresponding to a minimum halo mass $M_{\rm
  crit} \simeq 3 \times 10^6~\Ms$ that can support H$_2$
cooling\cite{Machacek_2001} and primordial star formation.  However,
the MMH (LWH) gravitationally collapse only after they reach the
atomic cooling limit at $z = 16.4 (15.3)$, having masses of $2.6 (5.8)
\times 10^7~\Ms$, an order of magnitude above $M_{\rm crit}$.  Upon
closer inspection, we find that both target haloes experience a period
of intense growth in mass (Fig \ref{fig:evo}a).  The MMH grows by a
factor of 30 over 30~Myr ($z = 21-19$) as it virializes.  The LWH
experiences two rapid growth events.  It first increases from $2
\times 10^5\Ms$ to $3 \times 10^6~\Ms$ between $z=19-18$ (15 Myr), at
which point its mass fluctuates just above $M_{\rm crit}$.  At
redshift $z=15.8$, it then dramatically grows by a factor of nine
within a span of 10~Myr.  Most of the accreted matter originates from
the parent filament, a major merger, and several minor halo mergers.
\change{The currently standard cold dark matter paradigm has this
  unique prediction of intense matter convergence below the atomic
  cooling limit and is not present in cosmologies that suppress power
  below this scale.}  Nevertheless it is rare, only occurring in
$\sim$$3 \times 10^{-4}$ of haloes (see Methods) with similar masses
and existing in an overdense large-scale environment.

Gas within these growing haloes is dynamically heated as it strives
for virial equilibrium, \change{whose heating rate is linearly
  proportional to the halo mass growth rate\cite{Yoshida_2003a}.
  Dynamical heating is only important when gas cooling is inefficient,
  particularly in rapidly growing low-mass haloes.  In combination
  with the LW negative feedback, it can further suppress any attempt
  at collapse.}  Both target haloes sustain substantial dynamical
heating during their rapid growth events, driven primarily through
major mergers.  We find that they are the dominant mechanism for
preventing Pop III star formation in these haloes.

The simulations follow the evolution of the target haloes until a
density of $10^{-15}~\mathrm{g\ cm}^{-3}$, at which point it is
certain that a collapsed object will form \change{(see Supplementary
  Videos)}.  Both haloes form a gravitationally unstable core with a
mass and radius of 30,000 (200,000) \Ms{} and 3 (15) pc for the MMH
(LWH), respectively.  The MMH grows gradually after its rapid
growth event, allowing the system to form a rotationally supported
disk that is comparatively cold at 300~K to the surrounding 10,000~K
gas.  The medium within the cold disk is turbulent, which causes
numerous weak shocks (Fig. \ref{fig:proj}d).  The disk then fragments
into three clumps (Fig. \ref{fig:proj}a, \ref{fig:proj}b), as thermal
and rotational support cannot counteract their gravitational forces
(see Methods), all of which proceed to collapse.  The morphology of
the LWH is completely different from the MMH because of a recent
major merger.  The collision causes a sheet-like overdensity
(Fig. \ref{fig:proj}e) that cools to 300~K, becoming gravitationally
unstable to fragmentation.  A single clump fragments from the sheet
(Fig. \ref{fig:proj}f) and undergoes a catastrophic collapse.  All
clumps in both target haloes have masses around 1000 \Ms.

The radial profiles of the gas density (Fig. \ref{fig:radial}a)
generally follow a power-law with a slope of --2 that is expected for
an isothermal collapse, which can be translated into a gas mass
enclosed (Fig. \ref{fig:radial}b).  Deviations from a power-law
originate from the two other clumps in the MMH, seen as spikes
around 1 pc, and the sheet-like structure in the LWH, seen as an
inflection point around 1 pc.  The gas inside the Jeans mass (marked
with squares) becomes shielded from the extragalactic LW background,
allowing for the H$_2$ fraction to increase to $10^{-3}$, sufficient
to cool the gas down to 300~K (Fig. \ref{fig:radial}c).  Inside a
radial mass coordinate of 1000 \Ms, adiabatic compression heats the
gaseous core to 600--800~K.

The key indicator for SMS formation is rapid gas inflow onto the
gravitationally unstable core, not the overall Jeans mass.  It has
been shown that accretion rates over $\sim$$0.04 \Msyr$ onto a
nascent, central core will result in SMS formation\cite{Hosokawa_2013,
  Schleicher_2013}.  Their weak hydrogen ionizing luminosities cannot
reverse these strong gas flows\cite{Hosokawa_2016, Sakurai_2015}.  The
respective infall mass fluxes (Fig. \ref{fig:radial}d) at the Jeans
mass are 0.17 and 2.1 \Msyr{} for the MMH and LWH at the final
time, above the critical value for SMS formation.  This ample supply
of inflowing gas provides fuel for the clumps within the central
unstable object.  The infall rates onto the clumps (Extended Data
Fig. \ref{fig:clump-infall}a) are between 0.03 and 0.08 \Msyr{} at
their boundaries but increase rapidly to $\sim$0.5 \Msyr{} at a radial
mass coordinate of 10,000 \Ms, suggesting that cores continue to grow
rapidly after the final snapshot of our simulation for 1~Myr,
\change{similar to the typical SMS lifetime} (Extended Data
Fig. \ref{fig:clump-infall}b).  We thus conclude that the two target
haloes will host SMS formation, and subsequently, a direct collapse
black hole (DCBH) with an initial seed mass at least 1,000 \Ms{} and
perhaps up to 10,000 \Ms{} within 1~Myr.

Using the formation requirements previously discussed, we can estimate
(see Methods) the DCBH formation rate per comoving volume to be
$1.1_{-0.91}^{+1.7} \times 10^{-3}$ SMSs per comoving Mpc$^{-3}$ (68\%
confidence interval) that form through this new formation scenario in
overdense regions of the Universe.  Given that only 0.01--0.1\% of the
universe is in such an overdense region, the global number density of
DCBH formation is predicted to be $10^{-7} - 10^{-6}$ per comoving
Mpc$^{-3}$, 100--1000 times higher than other
estimates\cite{Hirano_2017}.

SMSs and thus DCBHs forming in rapidly growing haloes, as proposed
here, will be tens of kpc away from the large-scale overdensity.  They
will take hundreds of Myrs, a substantial fraction of the
Universe's age at $z > 6$, to fall into the nearby group of galaxies.
We predict that these DCBHs will evolve to form the population of
faint quasars observed at $z \sim 6$ (ref. \citen{Onoue_2017,
  Kim_2018}).  This population will be within the reach of the James
Webb Space Telescope that will provide stringent constraints on their
number densities, directly comparable to our results here.

\begin{addendum}

\item[Acknowledgements] JHW thanks Andrew Benson for assistance with
  the code {\sc Galacticus}.  JHW was supported by NSF awards
  AST-1614333 and OAC-1835213, NASA grant NNX17AG23G, and Hubble
  theory grant HST-AR-14326.  JR acknowledges support from the EU
  commission via the Marie Sk\l{}odowska-Curie Grant - ``SMARTSTARS''
  - grant number 699941.  BWO was supported in part by NSF awards
  PHY-1430152, AST-1514700, OAC-1835213, by NASA grants NNX12AC98G,
  NNX15AP39G, and by Hubble theory Grants HST-AR-13261.01-A and
  HST-AR-14315.001-A.  MLN was supported by NSF grants AST-1109243,
  AST-1615858, and OAC-1835213.  The simulation was performed on Blue
  Waters operated by the National Center for Supercomputing
  Applications (NCSA) with PRAC allocation support by the NSF (awards
  ACI-0832662, ACI-1238993, ACI-1514580). The subsequent analysis and
  the resimulations were performed with NSF's XSEDE allocation
  AST-120046 on the Stampede2 resource.  This research is part of the
  Blue Waters sustained-petascale computing project, which is
  supported by the NSF (awards OCI-0725070, ACI-1238993) and the state
  of Illinois. Blue Waters is a joint effort of the University of
  Illinois at Urbana-Champaign and its NCSA.  The freely available
  astrophysical analysis code \texttt{yt}\cite{YT} and plotting
  library matplotlib was used to construct numerous plots within this
  paper. Computations described in this work were performed using the
  publicly-available {\sc Enzo} code, which is the product of a
  collaborative effort of many independent scientists from numerous
  institutions around the world.

\item[Author contributions] JHW and JR conceived the idea, performed
  the analysis, and drafted the paper. The Renaissance Simulations
  were conducted by HX and JHW, and the resimulations of the target
  haloes were conducted by JHW.  BWO performed the Monte Carlo
  analysis for the number density estimate.  All authors contributed
  to the interpretation of the results and to the text of the final
  manuscript.

\item[Competing Interests] The authors declare no competing interests.

\textbf{Additional information}\\
\textbf{Extended data} is available in the online version of the
paper.\\
\textbf{Supplementary videos} are available in the online version of
the paper.\\
\textbf{Reprints and permissions information} is available at
\url{www.nature.com/reprint}.\\
\textbf{Correspondence and requests} for materials should be addressed
to JHW.
     
\end{addendum}


\renewcommand{\figurename}{{\bf Extended Data Figure}}
\setcounter{figure}{0} 

\begin{methods}
\label{Methods}

\hspace{1em} \textbf{Cosmological simulation of early galaxy formation.}  The
Renaissance Simulations were carried out using the open source
adaptive mesh refinement code {\sc Enzo}\cite{Enzo_2014}, a physics
rich, highly adaptive code that has been well tuned for high-redshift
structure formation simulations\cite{Abel_2002, OShea_2007b,
  Turk_2009, OShea_2015}. The Renaissance Simulations have been
well-detailed previously in the literature\cite{Xu_2013, Xu_2014,
  Chen_2014, Ahn_2015, OShea_2015, Xu_2016b, Xu_2016}, and here we
only summarize the simulation characteristics relevant to this study.
All of the Renaissance simulations were carried out in a comoving
volume of (40 Mpc)$^3$.  We set the cosmological parameters using the
7-year WMAP $\Lambda$CDM+SZ+LENS best fit\cite{Komatsu_2011}:
$\Omega_{\rm m} = 0.266, \Omega_\Lambda = 0.734, \Omega_{\rm b} =
0.0449, h = 0.71, \sigma_8 = 0.81$ and $n = 0.963$.  The simulations
were initially run until a redshift $z=6$ at relatively coarse
resolution with $512^3$ particles each with a mass of $1.7 \times
10^7~\Ms$.  Three regions of interest were then selected for
re-simulation at much higher refinement - namely a rare-peak region, a
normal region and a void region. The rare-peak region has a comoving
volume of 133.6~Mpc$^3$, whereas the normal and void regions each have
comoving volumes of 220.5~Mpc$^3$.

In this study, we focus on the rare-peak simulation, which was
selected by extracting the Lagrangian region centered on two $3 \times
10^{10}$ \Ms{} haloes at $z=6$, the most massive at that time. The
dimensions of the rare-peak region were set at $5.2 \times 7.0 \times
8.3$ Mpc$^3$. The simulation was re-initialized using
MUSIC\cite{Hahn_2011} with a further three nested grids centered on
the rare-peak region. This led to an effective resolution of 4096$^3$
and a dark matter particle resolution of $2.9 \times 10^4$ \Ms{}
within the highest refinement region. During the course of the
simulation further adaptive refinement is allowed within the
refinement zone (i.e. the Lagrangian region of the rare-peak) up to a
maximum level of 12 leading to a maximum spatial resolution of 19
comoving pc (1.2~proper pc at $z = 15$).  The simulation was
halted at a final redshift $z=15$ due to the high computational
expense. The halo mass function is well-resolved down to $2 \times
10^6~\Ms$ with 70 particles per halo\cite{Xu_2016}, and at the ending
redshift, the simulations contained 822 galaxies having at least 1,000
particles ($M_{\rm halo} \simeq 2.9 \times 10^7~\Ms$).  We follow the
ionization states of hydrogen and helium with a 9-species primordial
non-equilibrium chemistry and cooling network\cite{Abel_1997},
supplemented by metal-dependent cooling tables\cite{Smith_2009}.  Dark
matter halo catalogs and the associated merger trees were created with
{\sc Rockstar}\cite{Behroozi_2013a} and {\sc
  consistent-trees}\cite{Behroozi_2013b}, respectively.

\textbf{Star formation and feedback.} The Renaissance Simulations
include both self-consistent Pop III and metal-enriched star
formation at the maximum refinement level.  The simulation captures
star formation in haloes as small as $3 \times 10^6~\Ms$
(ref. \citen{Xu_2013}).  Pop III star formation occurs when a
cell meets all of the following criteria:
\begin{itemize}
  \setlength\itemsep{0em}
\item An overdensity of $5 \times 10^5 (\sim 10^3\ \rm{cm^{-3}\ at\ z} = 10)$
\item A converging gas flow ($\nabla \cdot \rm{v_{gas}} < 0$)
\item A molecular hydrogen fraction $\rm{f_{H2} > 5 \times 10^{-4}}$
\end{itemize}
These physical conditions are typical of collapsing metal-free
molecular clouds $\sim 10$ Myr before the birth of a Pop III
main sequence star\cite{Abel_2002}. In this scenario, each star
particle within the simulation represents a single star.  Population
III star formation occurs if the metallicity is less than $10^{-4}$ of
the solar fraction in the highest density cell with metal-enriched
star formation proceeding otherwise. Pop III star formation furthermore
requires that the H$_2$ fraction is greater than $5 \times 10^{-4}$. This takes
into account the fact that star formation should not proceed in
the presence of a strong Lyman-Werner (LW) radiation field.  The
functional form of the IMF and supernovae feedback are
calibrated\cite{Wise_2012b} against high-resolution Pop III
star formation simulations, stellar evolution models, and observations
and models of star formation in local molecular clouds.  Stellar
feedback uses the {\sc Moray} radiative transport
framework\cite{WiseAbel_2011} for ionizing photons. LW radiation that
dissociates H$_2$ is modeled using an optically thin, inverse square
law profile, centered on all star particles.  We do not include any
H$_2$ self-shielding, which only becomes important at high densities.
In particular, DCBH host halo candidates only shield themselves from a
background at scales below 3 pc\cite{Regan_2016a}, which is close to
our resolution limit.  A LW background radiation field is also
included to model radiation from stars which are not within the
simulation volume\cite{Wise_2012b} that is added to the spatially
varying LW radiation field created by stars inside the volume. In the
high density region of the rare-peak simulation the LW radiation from
stars dominates over the background.  Although we cannot follow
Pop III star formation in haloes below $3 \times 10^6~\Ms$, it
is suppressed by the LW background in such haloes\cite{Machacek_2001,
  Wise_2007b, OShea_2008} and also by
baryonic streaming velocities in certain regions\cite{Naoz_2013, Hirano_2017}.
Thus, we are confident that our simulation
follows the complete star formation history of all collapsed
structure, and thus the metal-enrichment history of pre-galactic gas
that is vital to determine the conditions for DCBH formation.

\textbf{Direct-collapse black holes.} The star formation and feedback
models do not include a DCBH formation model but consider the
appropriate astrophysical processes to ascertain the chemical and
thermal state of all collapsed objects resolved by the simulation,
essential for searching for candidate DCBH formation sites.  Also
critical to any DCBH formation scenario is the inclusion of Population
III star formation and its supernova feedback that generally enriches
typical pre-galactic material with heavy elements.  The emergence of a
DCBH/SMS environment is therefore a robust prediction of the
Renaissance simulations.  The dark matter resolution in the original
Renaissance Simulations is not sufficient to follow the detailed
collapse of these objects\cite{Regan_2015}, however the resimulations
(see next section) of the target haloes have an ultimate mass and
spatial resolution of 103~\Ms{} and 60~AU, allowing us to accurately
follow the dynamics of the collapsing halo and determine whether their
gas infall rates are large enough to support SMS and thus DCBH
formation.

\textbf{High-resolution simulations of target haloes.} After
identifying the candidate haloes in the Renaissance Simulations, we
resimulate the two target haloes at higher mass and spatial resolution,
starting at a redshift $z = 20$.  We first identify the dark matter
particles within three virial radii of the target haloes at the final
redshift $z=15$.  Using their unique particle identifiers, we
determine their positions at $z=20$ and split their mass equally into
13 particles.  Twelve of child particles are placed at the 12 vertices
of a hexagonal close packed array, and the remaining particle is placed
at the original particle position\cite{Kitsionas_2002}.  We recursively
split the particles twice, decreasing the dark matter particle mass by
a factor of $13^2$ to $103~\Ms$ in the Lagrangian region of the target
haloes.  This method of particle splitting is widely used in
high-resolution cosmological simulations\cite{Bromm_2003, Dotti_2007,
  Hirano_2014, Regan_2015}, however this method may induce artificial
smoothing of the density field\cite{Chiaki_2015} and does not add
additional small-scale power to the matter distribution.

The Renaissance Simulations show that the target haloes remain
metal-free and are not exposed to any ionizing radiation.  Their
thermodynamic evolution and collapse are thus primarily controlled by
the halo growth history and the impinging LW flux.  With these priors,
we can safely neglect star formation and feedback, focusing on the
dynamics of their collapse.  We impose a uniform LW radiation
background that is time-dependent and is taken from the flux measured
at the center of the most massive progenitor halo in the original
simulation (see Fig. \ref{fig:evo}b).  Spatial deviations from this
value are less than a few percent within the Lagrangian region and
does not affect the thermodynamics of the gas.

In addition to the increased mass resolution, we increase the spatial
resolution by resolving the local Jeans length by at least 16 cells,
whereas the original simulation did not have enough mass resolution to
enforce such a refinement criterion.  We use this initial resimulation
to identify the time of collapse.  We then restart the simulations a
free-fall time (approximately 20~Myr) before the collapse time,
further increasing the spatial resolution so that the Jeans length is
resolved by at least 64 cells.  Computational limitations restrict us
from running at such high resolution for the entirety of the
resimulation, thus we only increase the resolution during the final
stages of the collapse.  We also include H$_2$ self-shielding of the
LW radiation field during this final
resimulation\cite{Wolcott-Green_2011} and compute the primordial
cooling rates with the software library {\sc Grackle}\cite{Grackle}.
We enforce a maximum refinement level of 24, corresponding to a
comoving (physical at $z=15$) resolution of 960~AU (60~AU).  We stop
the resimulations once they reach this maximum refinement level.  We
smooth the dark matter density field at scales below 9.5 comoving
pc (0.6 physical pc at $z=15$; refinement level 13).  At
these scales, the gas density dominates the matter density, and by
smoothing the dark matter density, we remove any artifacts associated
with the discrete representation of the dark matter mass
distribution\cite{Abel_2002}.

\begin{figure}[t]
  \centering
  \includegraphics[width=\columnwidth]{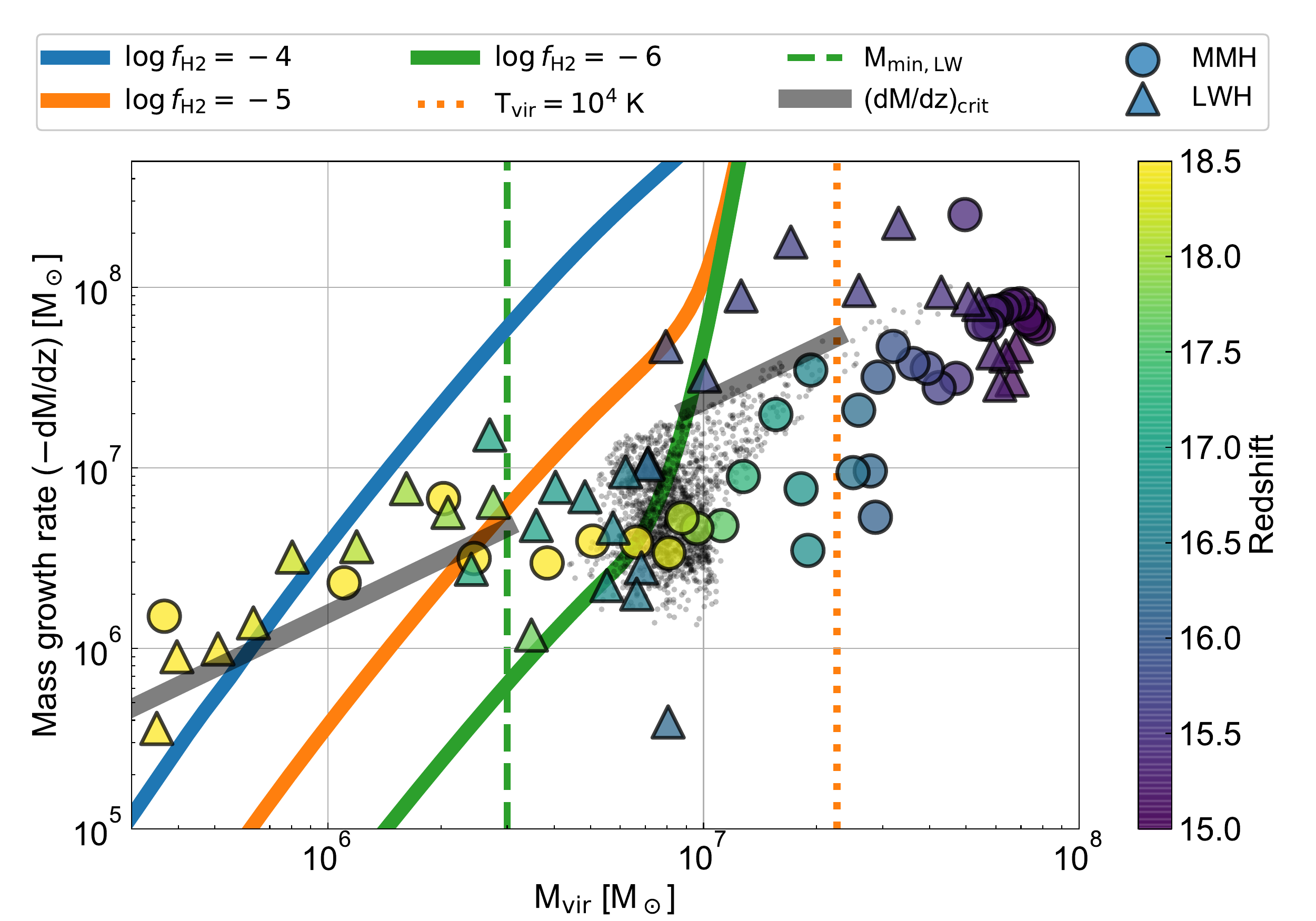}
  \caption{\textbf{Simulated and critical halo mass growth rates for
      SMS formation.}  A halo conducive for SMS formation must grow to
    the atomic cooling limit ($2.2 \times 10^7~\Ms$ at redshift 15;
    orange dotted line) without hosting star formation or being
    chemically enriched from nearby galaxies.  Haloes with masses
    below minimum mass $M_{\rm min,LW}$ (dashed green line) are
    suppressed by an external Lyman-Werner radiation field.  Above
    this mass, haloes with sufficient dynamical heating to suppress
    radiative cooling grow above a critical rate (Eq. \ref{eqn:crit}),
    which are shown for H$_2$ number fractions of $10^{-4}$ (blue
    solid line), $10^{-5}$ (orange solid line), and $10^{-6}$ (green
    solid line).  The simulated growth rates of the MMH (circles)
    and LWH (triangles) are above the $10^{-6}$ rate once it passes
    $M_{\rm min,LW}$. Above a halo mass of $8 \times 10^6~\Ms$ (a
    virial temperature of 8000~K at redshift 15), dynamical heating
    will not suppress cooling as the atomic radiative cooling rates
    are several orders of magnitude higher than molecular ones.  Both
    halos rapidly grow to $M_{\rm min,LW}$ that causes dynamical
    heating, preventing collapse for a sound-crossing time.  The
    LWH grows from $8 \times 10^6~\Ms$ to the atomic cooling limit
    within a dynamical time of the central core. Both conditions set a
    critical growth rate (thick solid gray lines). All other atomic
    cooling haloes (gray points) have similar growth rates between
    halo masses of $M_{\rm min,LW}$ and $8 \times 10^6~\Ms$ but far
    short of this critical growth rate.  Nearly all of these haloes
    cool and form stars before reaching the atomic cooling threshold.}
  \label{fig:growth}
\end{figure}

\textbf{Dynamical heating.} For a halo to potentially host a SMS/DCBH,
it must grow to the atomic cooling limit without forming stars or
being chemically enriched, thus any efficient cooling must be
suppressed.  LW radiation can suppress H$_2$ formation, and dynamical
heating can counteract any H$_2$ cooling in low-mass haloes.
\moved{The dynamical heating rate\cite{Yoshida_2003a} is given by
\begin{equation}
\Gamma_{\rm dyn} = \alpha M^{-1/3}_{\rm halo} \frac{k_{\rm
    B}}{\gamma-1} \frac{dM_{\rm halo}}{dt},
\end{equation}
where $\alpha$ is a coefficient\cite{Barkana_2001} relating the halo
virial mass and temperature ($T_{\rm halo} = \alpha M_{\rm
  halo}^{2/3}$), $M_{\rm halo}$ is the total halo mass, $k_{\rm B}$ is
the Boltzmann constant, and $\gamma = 5/3$ is the adiabatic index.
This process is only relevant when radiative cooling rates
$\Lambda(T_{\rm halo})$ per hydrogen atom have similar values.  Below
the atomic cooling limit, chemically primordial haloes rely on the
inefficient coolant H$_2$ to collapse and form stars.  Radiative
cooling will thus be suppressed when $\Gamma_{\rm dyn} > n\Lambda$,
where $n$ is the hydrogen number density.  Given this inequality, this
sets a critical halo growth rate\cite{Yoshida_2003a}
\begin{equation}
  \label{eqn:crit}
  \left(\frac{dM_{\rm halo}}{dt}\right)_{\rm crit} = \frac{2\alpha}{3}
  \frac{\gamma - 1}{k_{\rm B}} \, M_{\rm halo}^{-1/3} \,
  n\Lambda(T_{\rm halo})
\end{equation}
above which cooling is suppressed.}  Extended Data
Fig. \ref{fig:growth} compares the critical growth rates (colored
lines) for three different H$_2$ fractions, using Eq. \ref{eqn:crit},
and the growth rates of the two target haloes (circles and triangles).
We calculate the primordial cooling rate $\Lambda(T)$ for each H$_2$
fraction with {\sc Grackle}\cite{Grackle}.  Typical H$_2$ number
fractions $f_{\rm H2}$ in the target halo centers prior to collapse
are between $10^{-5}$ and $10^{-6}$.

\begin{figure}[t]
  \centering
  \includegraphics[width=\columnwidth]{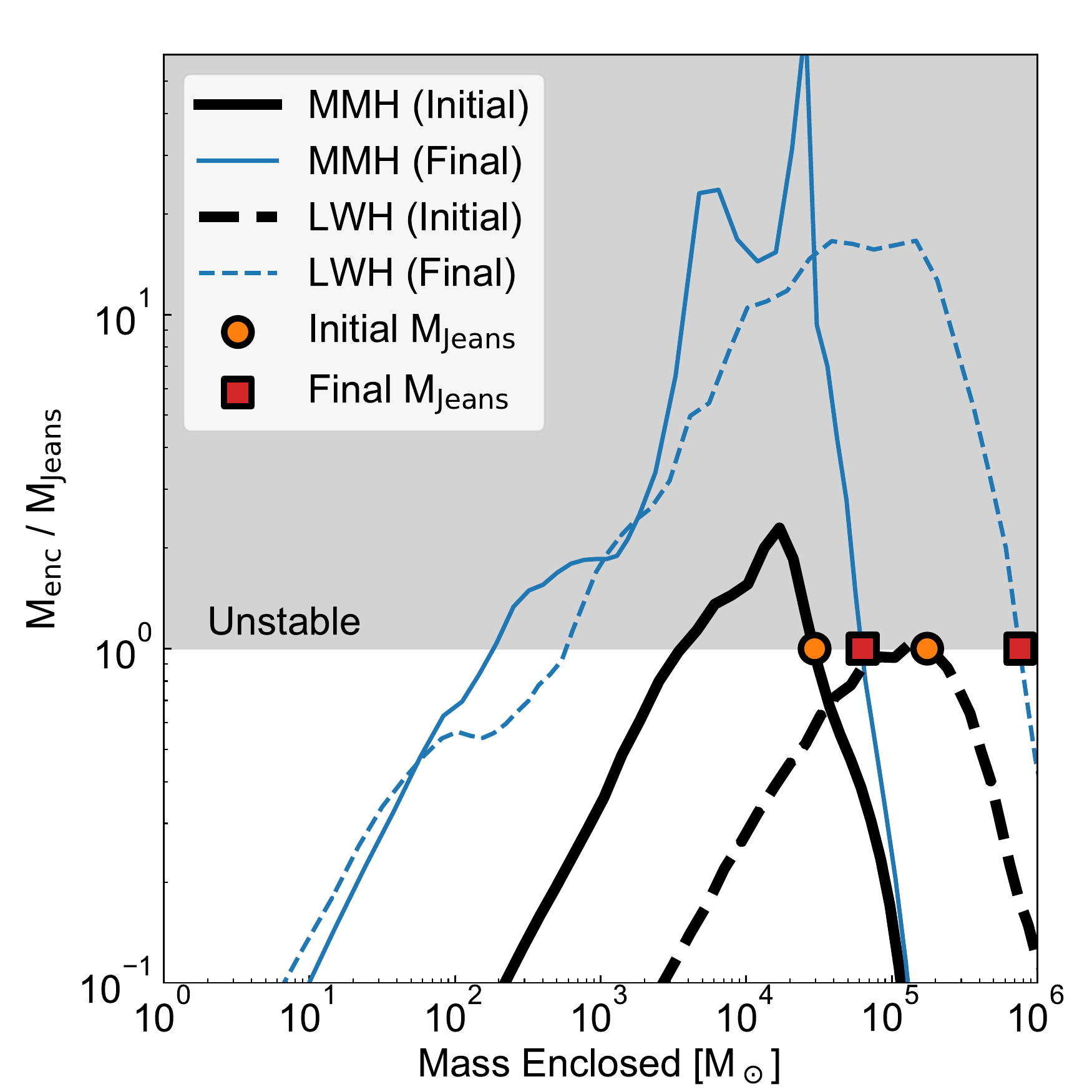}
  \caption{\textbf{Gravitational instability of the growing core.}
    The ratio of the enclosed gas mass $M_{\rm enc}$ and the Jeans
    mass $M_{\rm J}$ as a function enclosed gas mass is shown when the
    halo first becomes gravitationally unstable (thick black lines)
    when $M_{\rm enc} / M_{\rm J} \ge 1$ (shaded region) and the final
    simulation state (thin blue lines) for the MMH
    (solid lines) and LWH (dashed lines).  The orange
    circles and red squares indicate the mass scale of the collapsing
    gas cloud that is co-located with the center of the host halo.}
  \label{fig:jeans}
\end{figure}

Absent of any dynamical heating, haloes can cool and collapse through
H$_2$ once they reach a critical mass\cite{Machacek_2001, OShea_2008}
\begin{equation}
  \label{eq:mcrit_h2}
  \frac{M_{\rm min,LW}}{\Ms} = 1.25 \times 10^5 + 2.65 \times 10^5
  \left( \frac{J_{\rm LW}} {J_{21}} \right)^{0.47},
\end{equation}
shown as the dashed vertical line for $J_{\rm LW} = 1~J_{21}$, the
intensity at both target haloes when they have such a mass.  Atomic
hydrogen cooling becomes dominant over molecular cooling at $T \simeq
7000$~K (corresponding to a halo mass of $6 \times 10^6~\Ms$ at
$z=15$), indicated by the sharp rise in the critical curves, after
which dynamical heating becomes unimportant in the target haloes.  In
any chemically pristine halo unaffected by ionizing radiation,
gravitational collapse ensues above this mass within a free-fall
timescale.  Models of SMS formation require infall rates above
$\sim$0.04 \Msyr{} that are only driven by deep gravitational
potentials of atomic cooling haloes, whose mass is shown as the dotted
vertical line at a mass of $2.2 \times 10^7~\Ms{} [(1+z)/16]^{-3/2}$.

Both the MMH and LWH rapidly assemble as it virializes,
initially growing at a rate (-dM/dz) of $(1-3) \times 10^{6}~\Ms$ (per
unit redshift) when it has a mass $M \le 10^6~\Ms$.  Although H$_2$
formation is photo-suppressed at these low masses, dynamical heating
is present in these early rapid periods of growth and have lasting
effects on the halo gas.  The infalling gas shocks near the halo
center, heating the halo to the virial temperature.  Only after a
sound-crossing time, the halo comes into virial equilibrium.  This
takes 30~Myr for a $3 \times 10^6~\Ms$ halo (150 pc
radius) and a sound speed of 10 km/s.

As the haloes equilibrate, they continue to grow but at a reduced
rate, however still above the critical $f_{\rm H2} = 10^{-6}$ curve
until a mass of $6 \times 10^6~\Ms$.  The MMH accretes at a near
constant rate of $3 \times 10^6~\Ms$ (per unit redshift) with a peak
rate of $7 \times 10^6~\Ms$ (per unit redshift) at $M = 2 \times
10^6~\Ms$.  The LWH growth rate fluctuates between $10^6$ and
$10^7~\Ms$ (per unit redshift).  When atomic cooling becomes
efficient, catastrophic collapse occurs within a free-fall time
$t_{\rm ff} = \sqrt{3\pi / 32G\rho}$ that is 20~Myr for a
density of $10^{-23}$~g cm$^{-3}$, typical of haloes compressed
adiabatically\cite{Wise_2007a}.

As shown in Fig. \ref{fig:evo}, the MMH undergoes one rapid growth
event, starting at $2 \times 10^5~\Ms$ at $z = 21$ \moved{as it
  viralizes and occuring before H$_2$ cooling becomes efficient (and
  hence before the local LW flux becomes important).  The rapid infall
  creates a shock near its center, heating the halo gas over a
  sound-crossing time of 30~Myr.  The halo gas stabilizes after this
  event, and then it collapses on a free-fall time $t_{\rm ff} =
  \sqrt{3\pi/32G\rho} \simeq 20$~Myr, where $\rho \simeq 10^{-23}\,
  \textrm{g\ cm}^{-3}$ is the maximum gas density before collapse.}
After the rapid growth halts, the halo must grow to the atomic cooling
limit within 50~Myr, the sum of the sound-crossing time and free-fall
time.  Otherwise, it will collapse when atomic cooling is not
efficient, and it will cool through H$_2$ and fragment into more
typical Population III star formation.  It collapses at $z = 16.4$
when its mass is just above the atomic cooling limit.  There exists
75~Myr between these two redshifts, leaving 25~Myr for early rapid
growth.  Given that halo masses increase
exponentially\cite{Wechsler_2002}, i.e. $dM/dz \propto \alpha M$ with
$M(z) \propto e^{\alpha z}$, we can estimate a critical growth rate in
those first 25~Myr, shown as a left thick gray line in Extended Data
Fig. \ref{fig:growth}, using the mass difference between $z=21-19.2$
(25~Myr) when it grows to $3 \times 10^6~\Ms$.  The LWH experiences
two rapid growth episodes, where the first one is similar to the
MMH.  \moved{Afterwards, its mass hovers just above the critical
  value $M_{\rm crit}$ for efficient H$_2$ cooling.}  The later growth
spurt occurs just before it starts to cool atomically, seen by the
increase in (-dM/dz) to greater than $3 \times 10^7~\Ms$.  The
associated dynamical heating delays the collapse of the LWH until
its mass is $5.8 \times 10^7~\Ms$, twice the atomic cooling limit.

In general for dynamical heating to suppress metal-free star
formation, halo growth must be rapid from the H$_2$ cooling limit to
the atomic cooling limit.  More specifically, it must happen faster
than a free-fall time (20~Myr).  This sets another, more
general, critical growth rate between these two halo mass regions,
using the same approach as before.  It is shown as the thick gray line
between the green $f_{\rm H2} = 10^{-6}$ critical curve and the atomic
cooling limit.  Any halo that grows faster than this rate will have
its gravitational potential deepen faster than it can collapse, making
it more likely to support strong radial inflows, conducive for SMS
formation.

\textbf{Gravitational collapse of the central core.} After reaching
the atomic cooling threshold, the gas begins to cool rapidly and
(dynamical) heating mechanisms are suppressed.  Above a temperature of
approximately 8000~K, radiative cooling of neutral hydrogen sets an
upper limit to the temperature of the halo inside the virial radius,
inducing a rapid increase in density and ultimately gravitational
collapse. Extended Data Fig. \ref{fig:jeans} shows the ratio of the
enclosed mass, $M_{\rm enc}$, and the Jeans mass, $M_{\rm J}$, against
the enclosed gas mass. For gravitational collapse to take hold, the gas
inside the collapsing core must exceed the Jeans mass of the gas. In
Extended Data Fig. \ref{fig:jeans}, any lines entering the shaded
region are unstable to gravitational collapse. For the MMH, the
core of the halo first becomes unstable with an enclosed mass
approximately $3 \times 10^4$ \Ms{} -- marked by an orange circle.
As the halo grows in mass, the region grows to almost $10^5~\Ms{}$ by
the end of the simulation, which is gravitationally unstable and in
principal also subject to fragmentation. The LWH first crosses into
a region of instability with an enclosed mass of greater than
$10^5~\Ms$, and by the end of the simulation, the central region of
almost $10^6$ \Ms{} has become unstable to gravitational collapse.

\textbf{Support within the collapsing core.}  As the core regions of
both haloes start to undergo gravitational collapse, as outlined
above, both thermal, turbulent, and rotational support will act to
counteract the gravitational collapse. If there is sufficient gas
support, the gravitational collapse can be suppressed even though the
Jeans mass has been breached. Extended Data Fig. \ref{fig:velocities}
shows both the thermal and turbulent sound speeds as a function of
enclosed mass. The thermal sound speed is calculated in the usual way
as
\begin{equation}
  c_{\rm s} = \sqrt{{\gamma k_{\rm B} T} \over {\mu m_{\rm H}}},
\end{equation}
where $c_{\rm s}$ is the sound speed, $\gamma$ is the adiabatic index,
$k_{\rm B}$ is the Boltzmann constant, $T$ is the temperature, $\mu$
is the mean molecular weight and $m_{\rm H}$ is the hydrogen mass. In
calculating the total pressure support of the gas, the impact of the
turbulent velocity field must also be considered.  We calculate the
root mean squared turbulent velocities of the gas by subtracting the
bulk gas velocity from the velocity field.  We compute the bulk
velocity as the velocity of the native computational cells averaged
over a spherical $32^3$ grid that has cell widths much coarser than
the simulation data.  It has an outer radius equal to the virial
radius, equally spaced angular bins, and equally log-spaced radial
bins.  We then approximate the bulk velocity of each native cell with
the value of the spherical grid cell that contains the native cell
center.

\begin{figure}[t]
  \centering
  \includegraphics[width=\columnwidth]{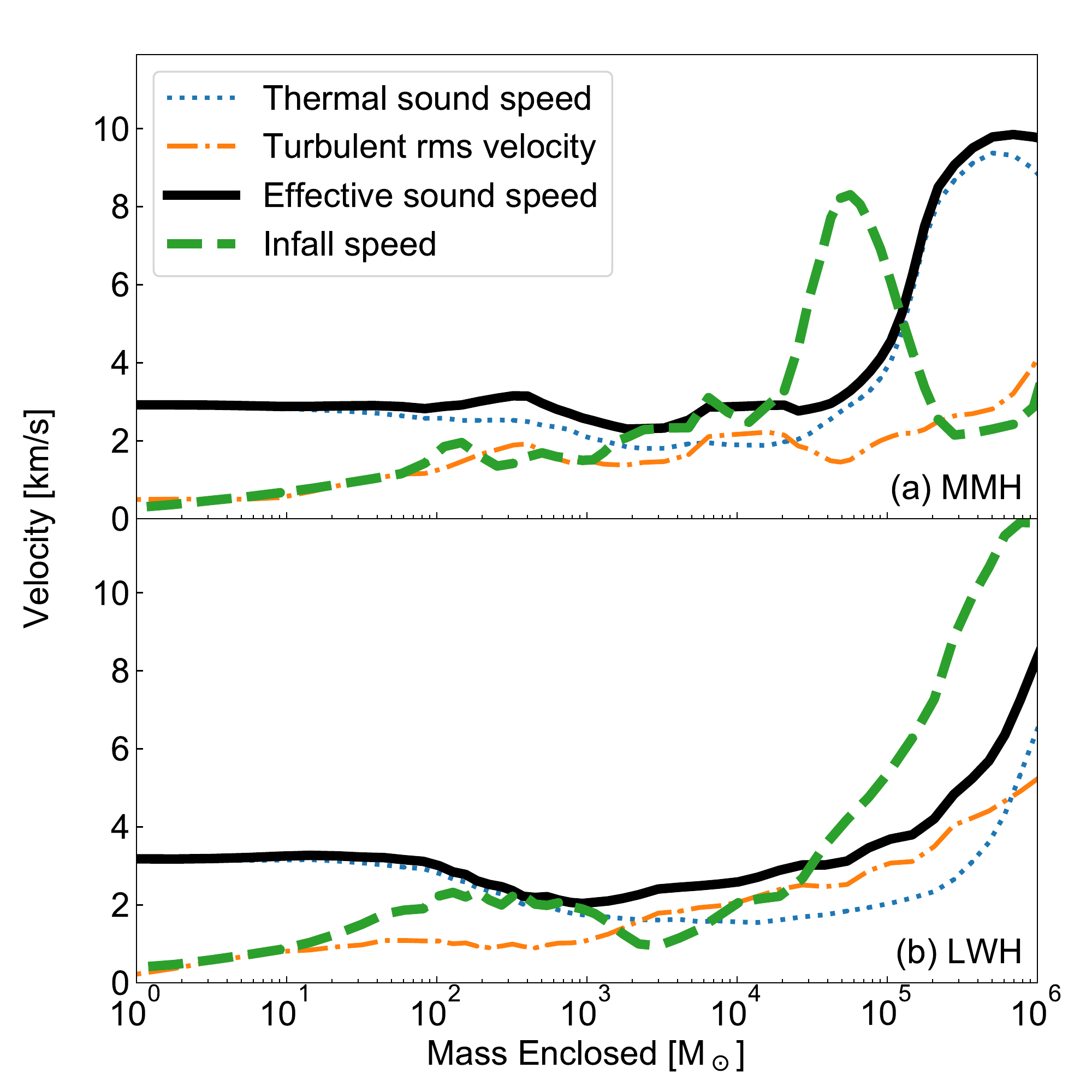}
  \caption{\textbf{Thermal and turbulent support of the collapsing
      core.} Gravitational forces dominate over thermal and turbulent
    internal pressures within the collapsing core in the MMH
    (\textbf{a}) and the LWH (\textbf{b}).  The
    thermal sound speed (blue dotted lines) and turbulent rms velocity
    (orange dash-dotted line) both contribute to the effective sound
    speed (black solid line) that provides partial resistance to a
    catastrophic collapse.  The radial infall speed (green dashed
    line) shows that the flow becomes supersonic at the Jeans mass
    scale and then transitions to a subsonic flow at smaller mass
    scales.  In the LWH, the radial inflow becomes
    transonic at a mass scale of 1000 solar masses.}
  \label{fig:velocities}
\end{figure}

The turbulent and thermal components act together to support the gas
against collapse and create an effective sound speed, $c_{\rm eff}^2 =
c_{\rm s}^2 + v_{\rm rms}^2$.  In Extended Data
Fig. \ref{fig:velocities}, we see that the infall speed always exceeds
the effective sound speed at some enclosed mass, indicating that, at
this scale, the thermal and turbulent support cannot support the gas
against gravitational collapse. For the MMH, the infall speed
exceeds the effective sound speed at approximately $10^5$
\Ms. This behavior is similar to the scale at which the gas also
becomes Jeans unstable, and so we define the collapsing core radius to
be equal to this scale.  For the LWH, the gas provides little or no
thermal or turbulent support, and the gas is free to collapse on
approximately the freefall timescale.

\begin{figure}[t]
  \centering
  \includegraphics[width=\columnwidth]{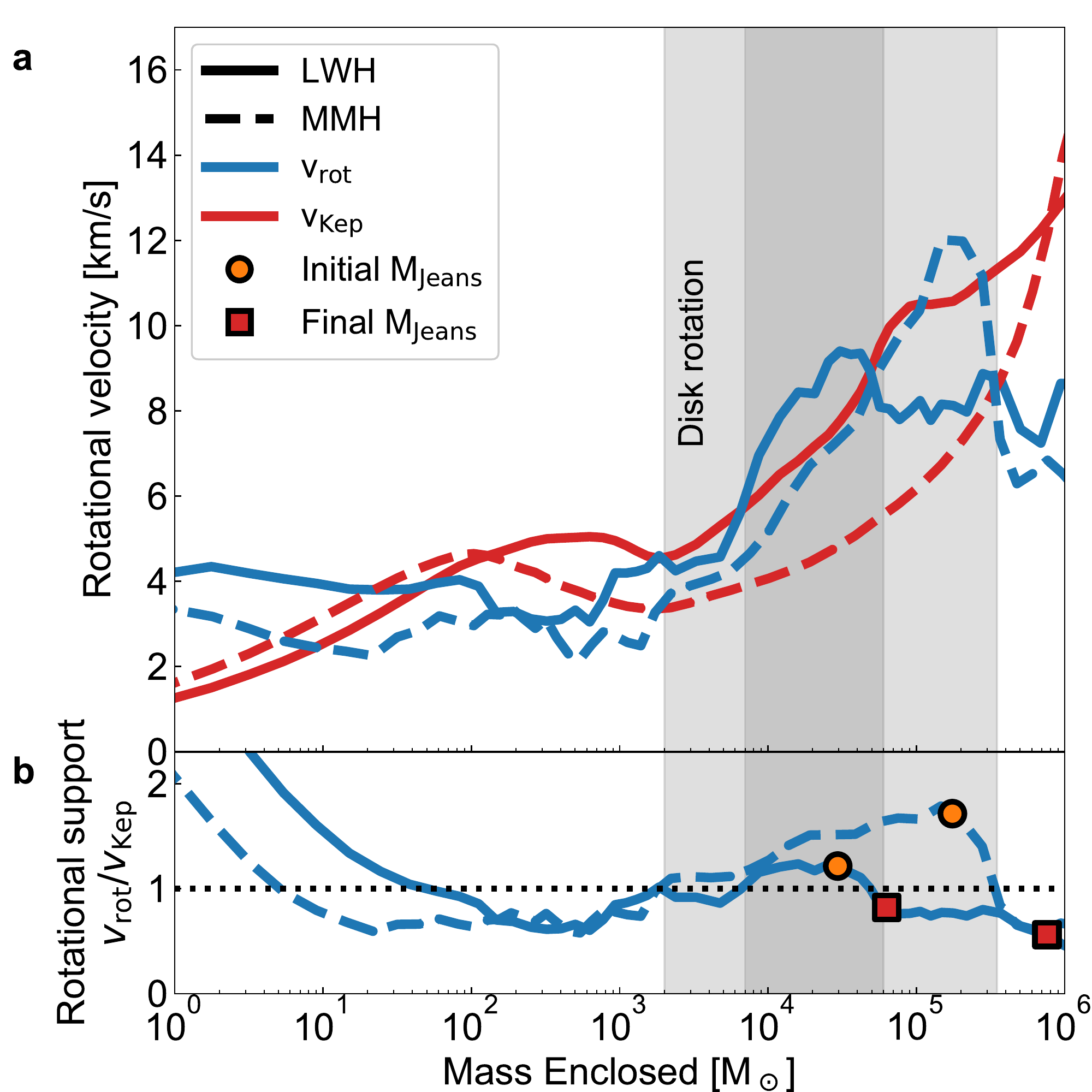}
  \caption{\textbf{Rotational properties of the target haloes.}  {\bf
      a}, Radially averaged profiles of circular velocity $v_{\rm Kep}
    = \sqrt{GM/r}$ (red lines) and rotational velocity $v_{\rm rot}$
    (blue lines) around the largest principal axis of the MMH
    (dashed lines) and LWH (solid lines) at the end of the
    simulation. {\bf b}, Radially averaged profiles of the fractional
    rotational support, where a ratio greater than one indicates
    rotational velocities are sufficient to prevent gravitational
    collapse.  The shaded regions show where the systems are
    rotationally supported, spanning from 2,000 to 330,000 (7,000 to
    60,000) solar masses for the MMH (LWH).  Rotation works in
    tandem with thermal and turbulent pressures to marginally slow the
    collapse, seen in the lower infall speeds at these mass scales in
    Extended Data Fig. \ref{fig:velocities}.  Inside 100 solar masses,
    this rotational measure becomes ill-defined because the rotation
    center and center-of-mass are not co-located, thus we do not
    conclude that the inner portions are rotationally supported even
    though $v_{\rm rot}/v_{\rm Kep} > 1$.}
  \label{fig:rotation}
\end{figure}

\begin{figure}[t]
  \centering
  \includegraphics[width=\columnwidth]{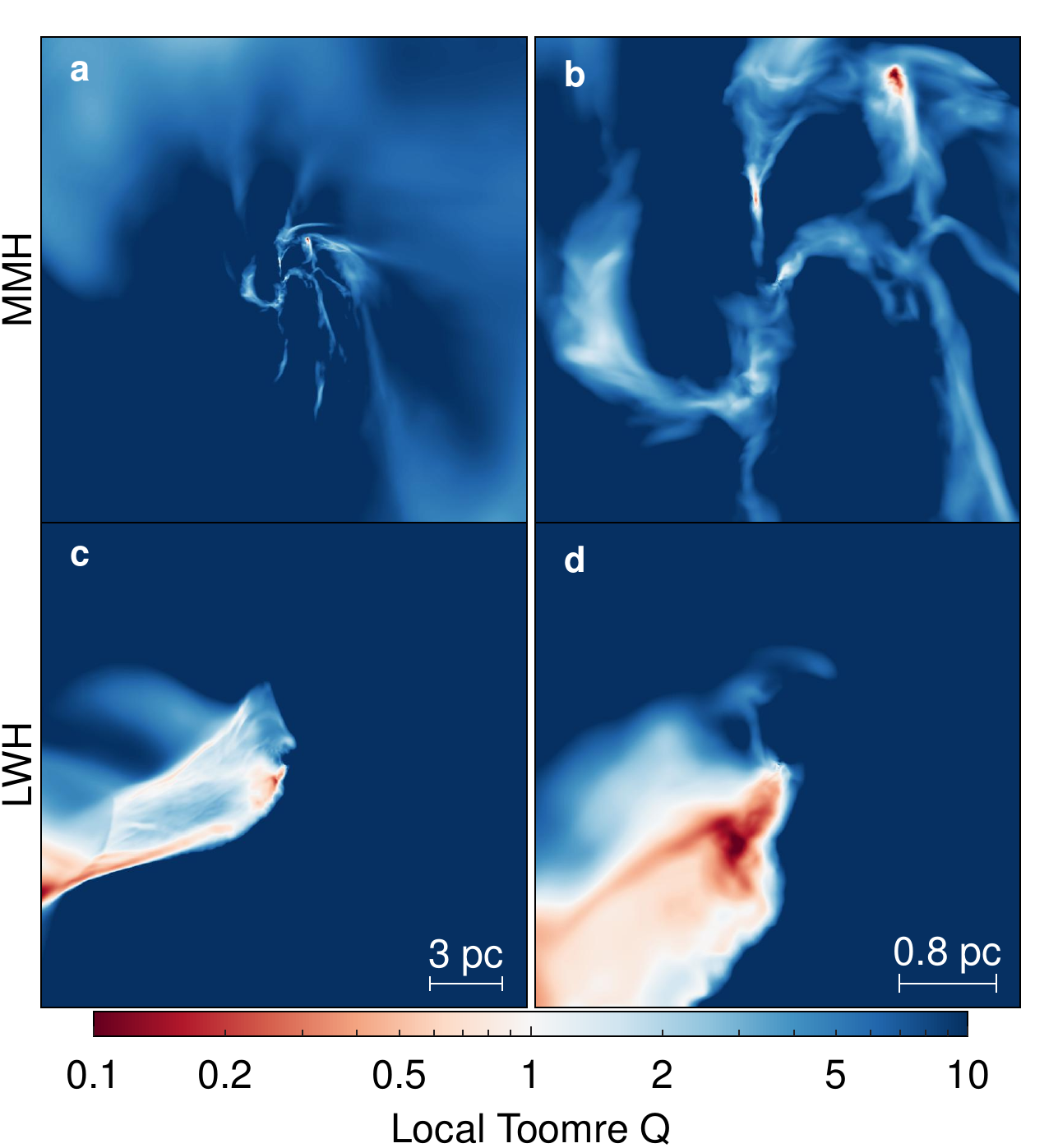}
  \caption{\textbf{Distribution of fragmentation-prone regions.}
    Density-weighted projections of a local estimate of the Toomre $Q$
    parameter (Eq. \ref{eqn:localQ}) for the MMH (\textbf{a,b}) and
    LWH (\textbf{c,d}) in a field of view of 20 (left) and 4
    (right) parsecs, centered on the densest point and aligned to be
    perpendicular with the angular momentum vector of the disk.  A
    value greater than one indicates that rotation and internal
    pressure stabilizes regions against fragmentation into smaller
    self-gravitating objects.  In the MMH, this analysis highlights
    the clump fragments with the filaments being only marginally
    stable at $Q \sim 1$.  The sheet in the LWH formed from a
    preceding major halo merger is apparent in this measure.  The bulk
    of the sheet is only marginally stable with the edge and
    collapsing center containing an environment conducive toward
    fragmentation.}
  \label{fig:toomre-proj}
\end{figure}

Interestingly in both cases, the radial inflow becomes transonic at
scales between $10^3$ \Ms{} and $10^4$ \Ms. This indicates that
at mass scales greater than approximately $10^3$ \Ms{} any fragments
are thermally and/or turbulently supported, and we do not expect
fragmentation of the gas cloud below this scale. Any protostars
forming within these clouds would have most of the gas contained in
the fragment available for accretion.  We will examine more closely
the possible fragmentation of the outer core into gas clouds in the
next subsection.

\begin{figure}[t]
  \centering
  \includegraphics[width=\columnwidth]{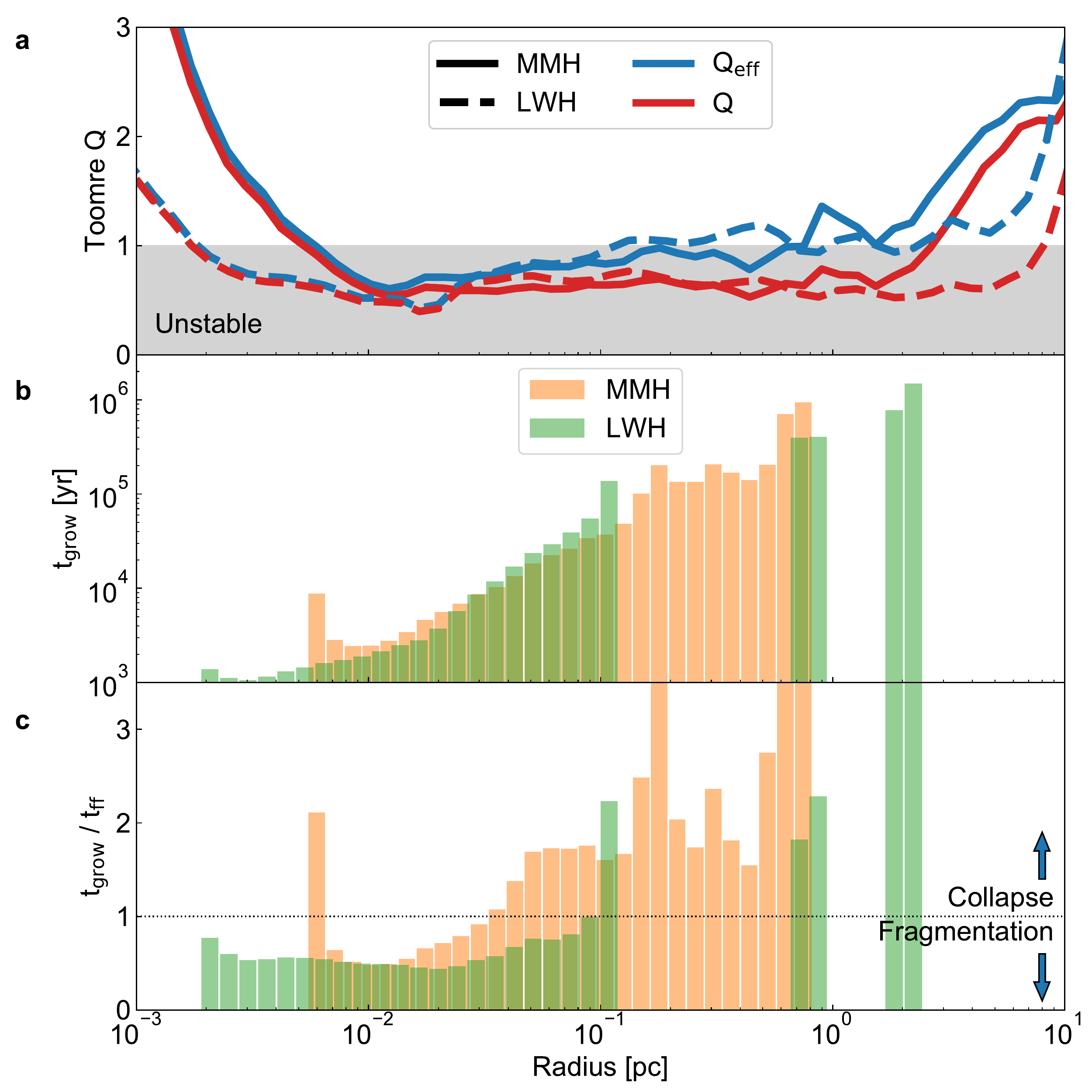}
  \caption{\textbf{Growth rates for fragmentation.}  A rotating system
    will only fragment into self-gravitating clumps when their growth
    rates are faster than the collapse timescale.  \textbf{a,}
    Cylindrical radial profiles of the Toomre Q parameter when
    considering only thermal support (red) and with thermal and
    turbulent support (blue) for the MMH (dashed) and LWH
    (solid).  The shaded region indicates where the system is unstable
    to fragmentation.  \textbf{b,} The unstable regions have a
    characteristic growth rate, defining a growth timescale $t_{\rm
      grow}$, showing an increasing trend with radius for the MMH
    (orange) and LWH (green). \textbf{c,} \change{If the ratio of $t_{\rm
      grow}$ and the free-fall time $t_{\rm ff}$ is less than one, the
    region can fragment before it gravitationally collapses.  In the
    MMH, this condition is true at radii less than 0.03 parsecs,
    indicating that small-scale fragmentation might occur but will be
    suppressed subsequently by a rapid monolithic collapse.  The LWH exhibits
    this feature inside 0.1 parsecs but is surrounded by gas that is
    stable against fragmentation.}}
  \label{fig:toomre}
\end{figure}

Extended Data Fig. \ref{fig:rotation} shows the rotational support
provided by the gas against gravitational collapse. The rotational
velocity of the gas is calculated by $|j|/a_1$, the ratio of the
specific angular momentum $|j|$ of the gas and the largest eigenvalue
$a_1$ of the inertia tensor, corresponding to the largest axis of the
system\cite{Regan_2009}.  The center of these profiles are taken to be
the maximum gas density, consistent with the rest of our analysis.  We
compare this rotational velocity against the Keplerian velocity
$v_{\rm Kep} = \sqrt{GM/r}$. The regions where the rotational velocity
exceeds the Keplerian velocity are shaded and indicate regions of
rotational support.  The rotationally supported region in the MMH
extends from 2,000 to 330,000 \Ms. For the LWH, the region of
rotational support is significantly smaller spanning only from 7,000
to 60,000 \Ms. The rotational velocities within 100~\Ms{} are not well
defined because the separation between the densest gas parcel and
rotation center becomes comparable to the radius (0.02 pc)
enclosing 100~\Ms{}.  We thus do not consider them to be rotationally
supported from this analysis.  The (disk-like) regions of rotational
support are nonetheless unstable to fragmentation and as we will see
fragmentation of the disk does occur in the MMH.

\textbf{Fragmentation of the collapsing core into gas clouds.}  As the
cloud collapses, cooling instabilities can cause the gas to fragment
into self-gravitating clumps.  Such fragmentation can be stabilized by
thermal pressure or centrifugal forces that can quantified by the
Toomre parameter,
\begin{equation}
  Q \equiv \frac{c_{\rm s} \kappa}{\pi G \Sigma}.
\end{equation}
The system is stable if $Q > 1$.  Here $\Sigma$ is the gas surface
density, and the epicyclic frequency $\kappa$ is directly calculated
from the data as
\begin{equation}
  \kappa^2 = \frac{2\Omega}{r}\frac{d(r^2\Omega)}{dr},
\end{equation}
where $\Omega = v_{\rm rot} / r$ is the angular velocity, and $r$ is
the radius in cylindrical coordinates with the $z$-axis aligned with
the total angular momentum vector of the Jeans unstable gas cloud.
These expressions assume an axisymmetric object.  However, the
collapsed objects in both haloes have non-ideal geometries.  It is thus
beneficial to consider a local measure of stability\cite{Hirano_2017}
\begin{equation}
  \label{eqn:localQ}
  Q_{\rm local} \simeq \frac{\Omega^2}{\pi G \rho}
\end{equation}
that can identify unstable regions with arbitrary geometries.
Extended Data Fig. \ref{fig:toomre-proj} shows $Q_{\rm local}$ for
both haloes in fields of view of 20 and 4 pc.  In the MMH, the
disk-like object is marginally stable in the spiral overdensities
(panel a).  Within those arms, there exists three clumps that have
fragmented (panel b) and started to collapse.  In the LWH, the
overdense sheet-like object, which is induced by a major merger, is
unstable to fragmentation in many regions and marginally stable in the
remainder of the object.  However, their collapse times are spread
out, resulting in only one (the first) clump to fragment out of the
overdense sheet by the end of the simulation.

Extended Data Fig. \ref{fig:toomre}a shows the Toomre Q parameter as a
function of cylindrical radius, where we have aligned the cylindrical
$z$-axis with the total angular momentum vector of the inner 10
pc.  We compare it with the effective Toomre Q parameter $Q_{\rm
  eff}$, where we substitute $c_{\rm s}$ with $c_{\rm eff}$ to include
any turbulent pressure support in the stability analysis.  When only
considering rotational support as a counterbalance to gravitational
collapse, the MMH and LWH are unstable ($Q < 1$) within a radius
of three and eight pc, respectively.  As we have shown earlier,
turbulent pressures are comparable to thermal pressures, and this
additional support stabilizes the rotating system against
fragmentation ($Q_{\rm eff} > 1$) outside 0.8 and 0.1 pc in the
MMH and LWH.  The LWH is also susceptible to fragmentation at
radii of 0.8 and 2 pc, where $Q_{\rm eff}$ becomes slightly less
than unity.

From the dispersion relation
\begin{equation}
  w_{\rm g}^2 = \kappa^2 + k^2 v_{\rm rot}^2 - 2\pi G k \Sigma
\end{equation}
that is used to calculate the growth perturbations in a gaseous disk,
the associated maximum growth rate\cite{Wang94} of a perturbation is
given by
\begin{equation}
  | w_{\rm max} | = \frac{\kappa ( 1 + Q_{\rm eff}^2 )^{1/2}}{Q_{\rm eff}},
\end{equation}
that occurs at $k_{\rm max} = \kappa / c_{\rm eff} Q$.  Here $k$ is
the wavenumber, and $w_{\rm g}$ is the growth rate.  We define the
growth time $t_{\rm grow} \equiv 1/ w_{\rm max}$ where $Q_{\rm eff} <
1$ (Extended Data Fig. \ref{fig:toomre}b).  For such a perturbation to
form before the system collapses, the growth time must be
\change{smaller} than a free-fall time (Extended Data
Fig. \ref{fig:toomre}c), i.e. $t_{\rm grow}/t_{\rm ff} < 1$.
\change{For the MMH, this condition is valid for radii smaller than
  0.03 pc, corresponding to 100~\Ms{} of enclosed gas mass.
  Between this radius and 1 pc, the system is unstable to
  fragmentation, but it is collapsing faster than it can fragment.
  This analysis is centered on the densest point, contained in one of
  the clumps.  This indicates that any fragmentation at small scales
  will be surrounded by a monolithic rapid collapse, most likely
  suppressing further fragmentation as this matter falls inward.  The
  two other clumps in the MMH form about 1 pc away.  Because we
  compute these quantities within cylindrical shells, the clumps with
  small $Q_{\rm local}$ (see Extended Data Fig. \ref{fig:toomre-proj})
  are averaged out.  Once they fragment, they begin to gravitationally
  collapse.  In the LWH, fragmentation can occur before collapse
  inside a radius of 0.1 pc.  Here a single clump fragments from
  the sheet-like overdensity, produced by a recent major merger, with
  a radius of $\sim$1 pc, which collapses faster than it can fragment
  at this length scale.}  This analysis shows that turbulence plays a
role in providing additional stabilization against fragmentation.  The
suppression of fragmentation allows for only a few clumps, at most, to
form in the collapsing system before the most dense clump collapses on
the order of its free-fall time, $t_{\rm ff} \sim 5{,}000$ yr.
This behavior allows for the rapid infall to be concentrated on these
few clumps, strengthening the case of SMS formation.

Given that the system is unstable to fragmentation, we search for
self-gravitating clouds with the clump finder provided with the
analysis toolkit {\tt yt}\cite{YT}.  We search for clumps that are
gravitationally bound and have a minimum of 20 cells, where clumps are
defined as topologically connected sets of cells and form a
hierarchical set of objects.  Because the LWH only has one clump
that contains the densest point, we only perform this analysis on the
MMH.  We identify three clumps that are also visible in the density
projections (Fig. \ref{fig:proj}b).  Below we have defined $t$ as the
time before collapse, where $t=0$ is a free-fall time after the final
simulation output of the maximum density.  The three clumps are
orbiting in the disk, located 1--2 pc from the rotational center,
where there is no local density maximum present.
\begin{itemize}
\item \textbf{Clump 1} contains the densest point in the simulation,
  and the projections (Fig. \ref{fig:proj}; Extended Data
  Fig. \ref{fig:toomre-proj}) and global radial profiles
  (Fig. \ref{fig:radial}; Extended Data Fig. \ref{fig:jeans},
  \ref{fig:velocities}, \ref{fig:rotation}, \ref{fig:toomre}) are
  centered on it. Its mass initially fluctuates between 200 and
  600~\Ms{} at $t = 30-70$~kyr as two clumps orbit each other,
  changing between topologically connected and disconnected in the
  process.  Their separation decays as energy is dissipated during
  their close encounters, eventually merging into a single clump at $t
  = 25$~kyr.  It grows to a mass of 790~\Ms{} and is located 1.2
  pc from the rotational center at the final time.
\item \textbf{Clump 2} is an elongated object that steadily grows in
  mass from 200 to 840~\Ms{}, orbiting at a distance of 1.8 pc,
  located above the Clump 1 in the projections.
\item \textbf{Clump 3} initially has a mass of 770~\Ms{}, orbiting at
  a distance of 2.2 pc, located in the upper-right of the
  projections.  At $t=50$~kyr, it merges with another clump,
  increasing its mass from 950 to 1400~\Ms{}.  The change in its
  distance from the rotation center is associated with the change in
  its center of mass after the merger.  It grows very little
  afterwards to a final mass of 1500~\Ms{}, while it migrates outward
  from 1.6 to 1.9 pc.
\end{itemize}
As the clumps evolve, they do not migrate inwards toward the disk
center, suggesting that mergers are not imminent between these three
clumps.  This strengthens the argument for 3 separate sites of SMS
formation and potentially DCBH formation. Such systems may be
detectable in the future through gravitational
waves\cite{Hartwig_2018}.

\begin{figure}[t]
  \centering
  \includegraphics[width=\columnwidth]{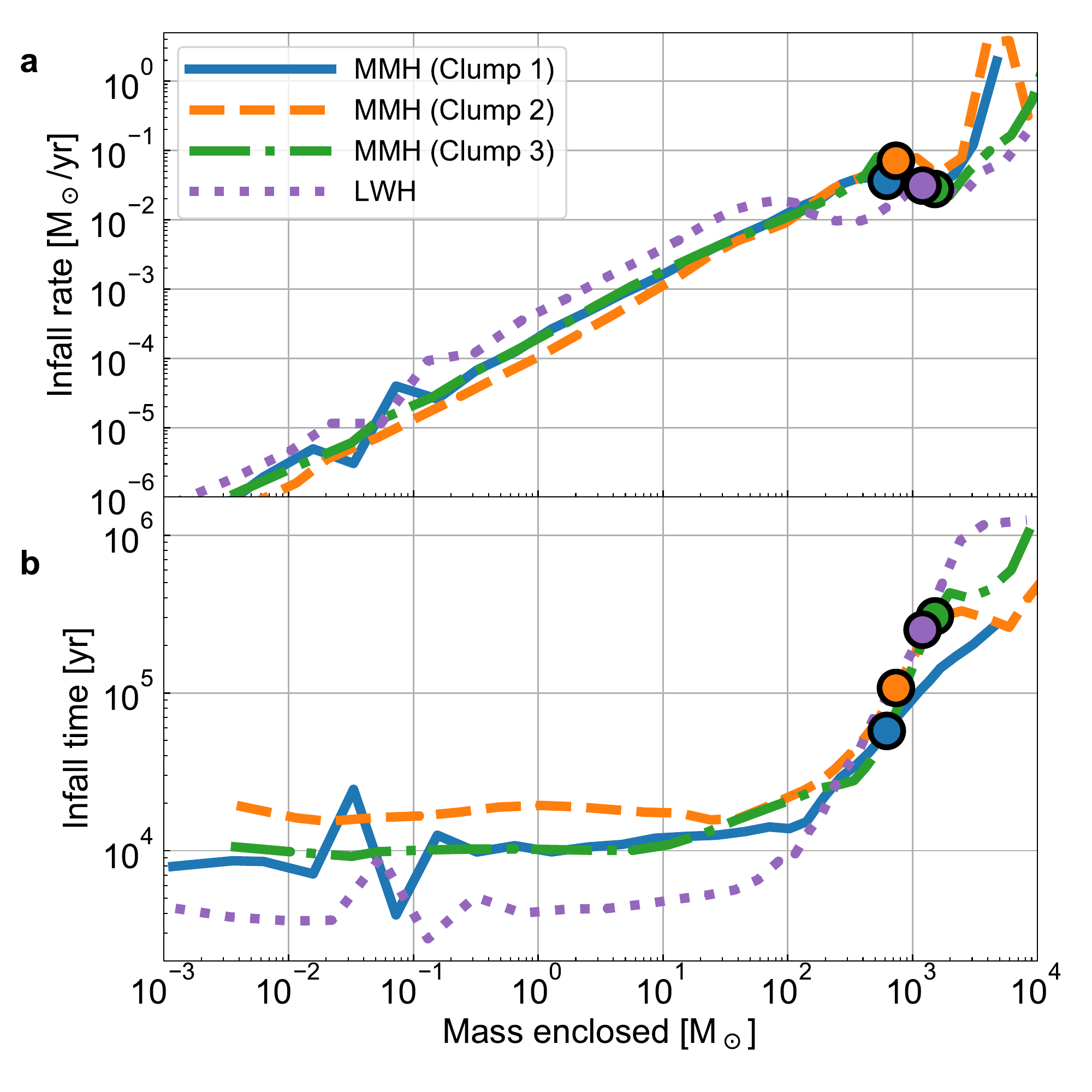}
  \caption{\textbf{Clump infall rates and timescales.} Similar to the
    results presented in Fig. \ref{fig:radial}d, the self-gravitating
    clumps are growing through radial infall.  \textbf{a,} The infall
    rates are computed as the mass flux through spherical shells,
    showing a steadily increasing rate with enclosed mass.  The rate
    in the single clump of the LWH (dotted purple
    line) is more than a factor three greater than the three major
    clumps in the MMH.  The circles mark the infall rate
    at the clump mass.  \textbf{b,} The infall time, the ratio of the
    mass enclosed and infall rate, is an informative scale that can be
    used to compare against star formation timescales.  This timescale
    is constant and approximately 10,000 years within 100 solar masses
    for all clumps and rises to $\sim$100,000 years for the entire
    clump, marked by the circles.  This rapid infall suggests that
    sufficient mass will collapse into supermassive protostar before
    it reaches main sequence.}
  \label{fig:clump-infall}
\end{figure}

\textbf{Collapse characteristics of the fragments.} We now examine
the accretion rates and thermal support of the clumps that form in
both the MMH and LWH. In Extended Data Fig.
\ref{fig:clump-infall}a, we plot the infall rate of the gas versus the
enclosed mass for each of the clumps (three for the MMH and one for
the LWH).  The infall rates are computed as the mass flux through
spherical shells and show a steady and near monotonic increase in
infall rate as a function of enclosed mass. The infall rate for the
LWH is more than a factor of three greater than the three major
clumps identified within the MMH. This is likely due to the
increased mass, at collapse time, and the higher radial velocity
associated with the LWH. The LWH experiences a major merger less
than 10 Myr before the end of the simulation that drives gas towards
the center thus explaining the increased infall-rates.

Circles in panel (a) mark the infall rate at the clump mass. This is
the infall rate that the outer part of the gas clump experiences. The
rates onto each clump are very similar with values between 0.01 and
0.1 \Msyr. These values are consistent with the values obtained for a
singular isothermal sphere collapse originally investigated by Shu
(1977)\cite{Shu_1977}. However, it should be noted that this is the
infall rate onto the outer gas clump and must be treated as an upper
limit when ascribing it to the accretion rate onto a protostar which
may subsequently form within the gas clump. Nonetheless, the infall
rates are very high and sufficient to drive a protostar towards
SMS formation.

In panel (b) we plot the infall time against enclosed mass, which is
calculated as the ratio of the mass enclosed and infall rate. It can
be used to directly compare against star formation timescales. Within
an enclosed mass of 100 \Ms{} the timescale for each clump is
approximately 10~kyr.  The Kelvin-Helmholtz timescale for massive
stars (M$_\star \lesssim 100$ \Ms) is less than 100~kyr. The
stars that form within these gas clumps will therefore reach the main
sequence while still accreting as is the case in SMS formation. Given
the timescales shown here, the infall rates suggest that the
(super)massive protostars will reach masses of at least 100 \Ms{}
before reaching the main sequence\cite{Bonnell_1998}. If the accretion
onto the clump continues to remain rapid at that point, full
gravitational contraction will be avoided (so-called ``hot accretion''
flows), and the stellar envelope will remain bloated, leading to SMS
formation in this context\cite{Hosokawa_2010, Hosokawa_2012,
  Hosokawa_2013, Woods_2017}.  \change{After its formation, SMS
  lifetimes are around 1~Myr, suggesting that the whole gas cloud
  could be accreted before the star exhausts its hydrogen supply.
  SMSs with accretion rates below 0.1 solar masses per year experience
  a general relativistic instability, creating a massive black hole
  with a mass similar to its progenitor mass, when its nuclear fuel is
  exhausted.  For higher accretion rates, the collapse occurs when the
  star is still burning hydrogen or helium\cite{Umeda_2016}, producing
  black holes with masses between 200,000 and 800,000~\Ms.  Lastly,
  there is a peculiar case when the SMS has a mass of around
  55,000~\Ms{} that produces an extremely energetic
  supernova\cite{Chen14}.}

\begin{figure}
  \centering
  \includegraphics[width=\columnwidth]{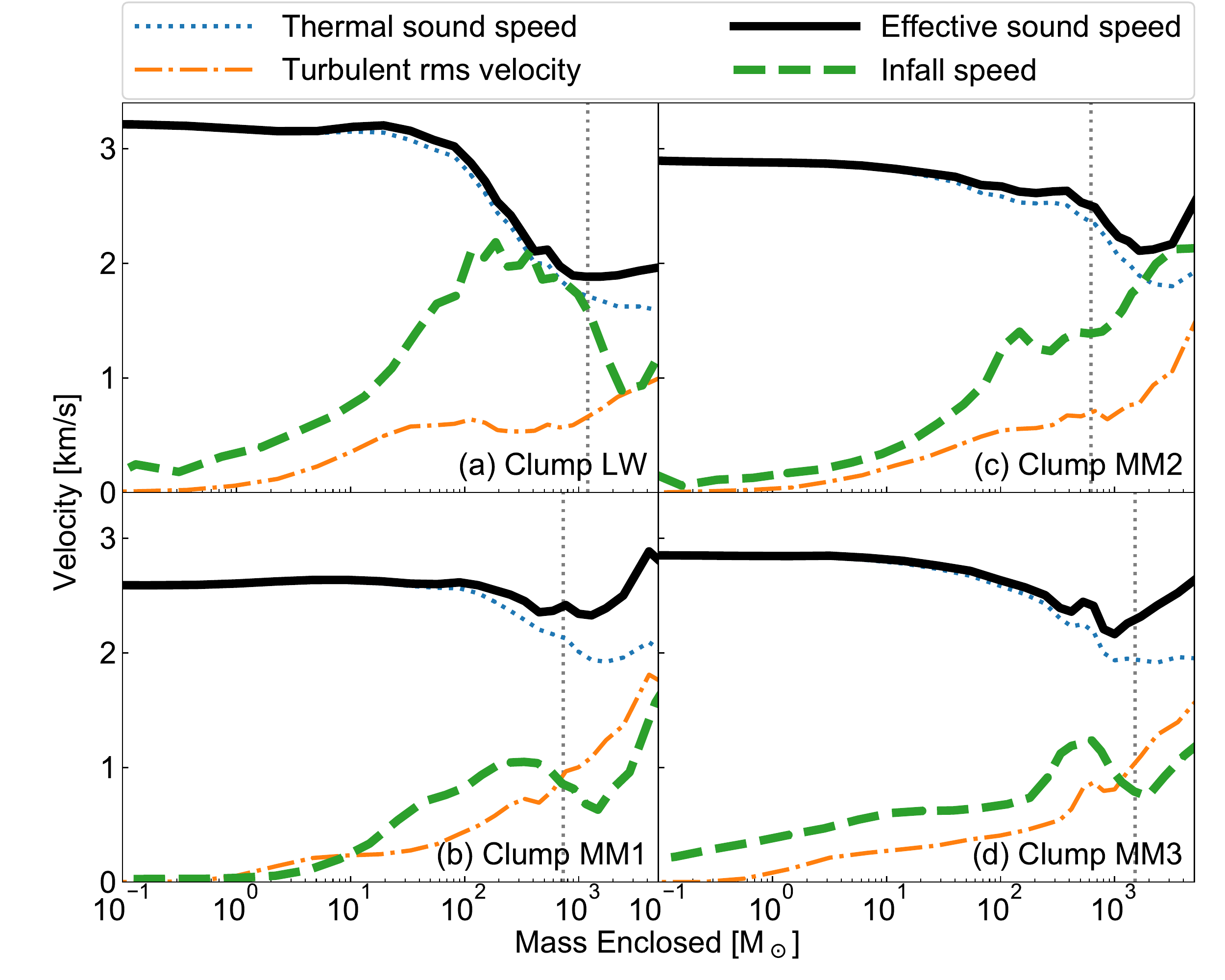}
  \caption{\textbf{Thermal and turbulent support of collapsing
      clumps.} Same as Extended Data Figure \ref{fig:velocities} but
    for the clumps in the LWH (\textbf{a}) and MMH
    (\textbf{b-d}).  The vertical dotted lines mark the
    clump mass in each panel.  The radial inflows are subsonic for all
    four clumps, however the clump in LWH contains
    transonic flows between 100 and 1,000 solar masses.  Thermal
    support is dominant inside the clumps, unlike the larger parent
    Jeans-unstable system where turbulent effective pressures are
    comparable to their thermal counterparts (see Extended Data Figure
    \ref{fig:velocities}).}
  \label{fig:clump-vel}
\end{figure}

In Extended Data Fig. \ref{fig:clump-vel}, we plot the fractional
support against collapse, similar to Extended Data
Fig. \ref{fig:velocities}. We do this by again comparing the thermal,
turbulent and infall velocities of the clumps.  Panel (a) shows the
velocities of the clump found for the LWH. Panels (b), (c), and (d)
show the clump velocities of each of the three clumps found for the
MMH. We find that in all cases the clumps are thermally supported
and stable against further gravitational collapse. The radial inflows
are subsonic for all four clumps, however the clump in the LWH
contains transonic flows between 100 \Ms{} and 1000 \Ms.  We
also note that the thermal pressure support is strongly dominant over
turbulent pressure support in all four clumps. This is in contrast to
the Jeans-unstable parent cloud where the turbulent pressures were
dominant.

\begin{figure}[t]
  \centering
  \includegraphics[width=\columnwidth]{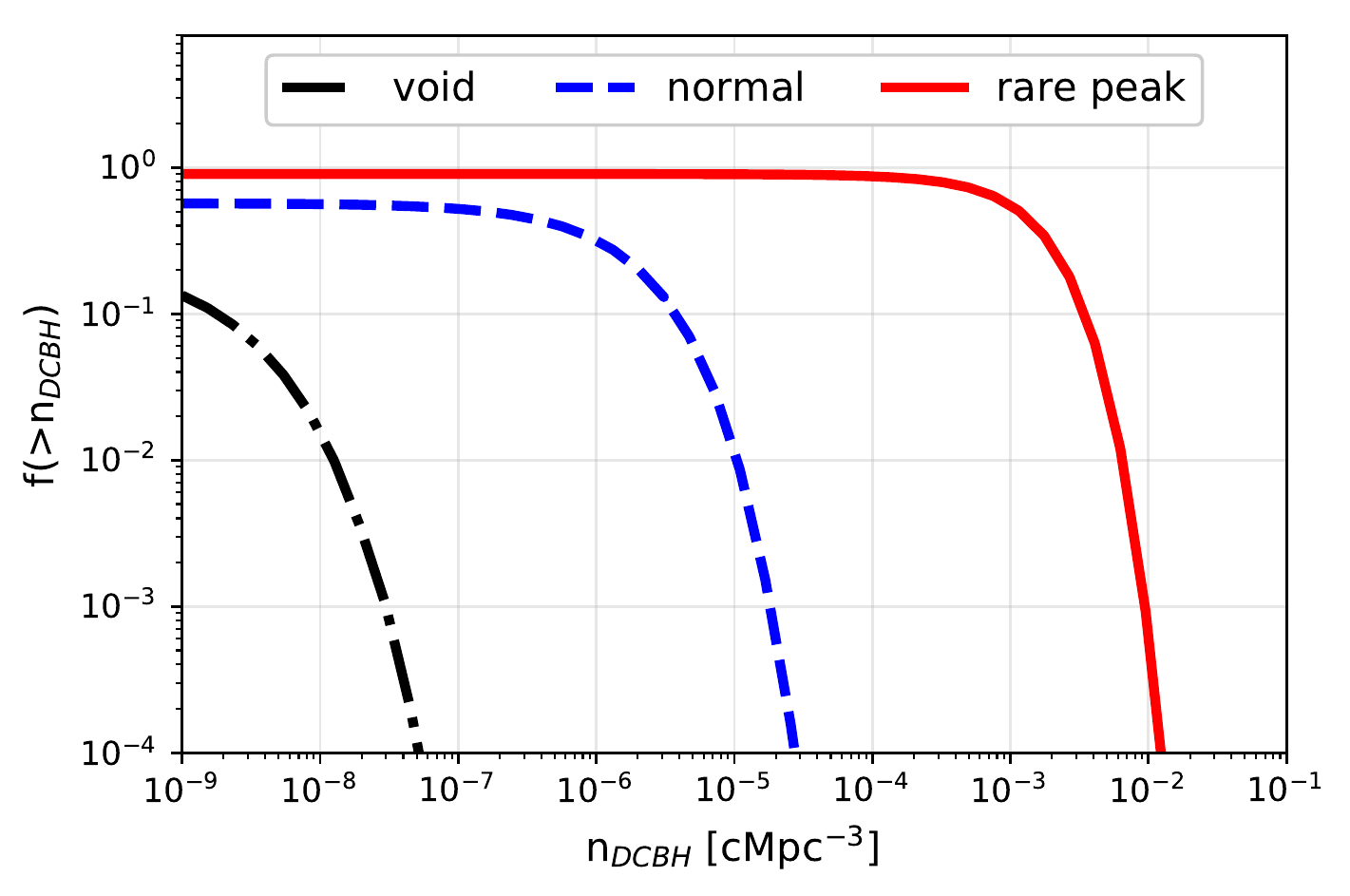}
  \caption{\textbf{Abundance estimate of direct collapse black holes.}
    The cumulative probability of the halo comoving number density
    that potentially host supermassive star formation shown for the
    rare-peak (red solid line), normal (blue dashed), and void (black
    dash-dotted) regions of the Renaissance Simulations.  Their
    respective median number densities are $1.1 \times 10^{-3}$, $\sim
    10^{-7}$, and 0 per comoving cubic megaparsec.  Subsequent direct
    collapse black hole formation are most likely to occur in
    overdense regions of the early universe, whereas little to no will
    form in average and underdense regions.}
  \label{fig:abundance}
\end{figure}

\textbf{Estimating the number density of SMS/DCBH formation sites.}
We estimate the comoving number density of DCBH formation sites,
${n_{\rm DCBH}}$, through the following expression: ${n_{\rm DCBH} =
  n_{\rm ACH} f_{\rm prim} f_{\rm LW} f_{\rm rapid}}$.  Here ${n_{\rm
    ACH}} = 5.0 \pm 0.19$ per comoving Mpc$^3$ is the comoving number
density of atomic-cooling haloes at redshift $z=15$ (i.e., haloes
above the mass threshold where cooling via atomic hydrogen is
effective); ${f_{\rm prim}} = 0.015 \pm 0.0045$ is the fraction of
those haloes that are of primordial composition; ${f_{\rm LW}} = 0.1
\pm 0.05$ is the fraction of the atomic cooling haloes that are in
regions where J$_{21} \geq 3$ (i.e., the fraction of the simulation
volume above that flux threshold); and ${f_{\rm rapid}} = 0.20 \pm
0.14$ is the fraction of the primordial atomic cooling haloes that are
growing above the critical threshold where accretion heating is
greater than cooling.  Based on the quantities found in the rare-peak
region of our simulation, we estimated these values and their
variance, which are given in the main article.  We then perform a
Monte Carlo sampling of the parameter space.  Extended Data
Fig. \ref{fig:abundance} shows the cumulative probability of the
expected DCBH number density in different large-scale environments.
They steeply decrease above $10^{-3}$, $10^{-6}$, and $10^{-9}$ DCBHs
per comoving cubic megaparsec in the rare-peak, normal, and void
regions of the Renaissance simulations.  The rare-peak has a median
value of $n_{\rm DCBH} = 1.1 \times 10^{-3}$ haloes/cMpc$^3$, a 68\%
confidence interval of $1.9 \times 10^{-4} - 2.8 \times 10^{-3}$
haloes/cMpc$^3$, and a 95\% confidence interval of $0 - 5.2 \times
10^{-3}$ haloes/cMpc$^3$ in the rare-peak region, where ``cMpc''
denotes comoving Mpc.

We note that the values described above represent the rare-peak region
only.  When the same quantities are extracted from the normal and void
simulation regions at $z=15$, the median number density of DCBH haloes
is expected to be $\sim 10^{-7}$ haloes/cMpc$^3$ and 0
haloes/cMpc$^3$, respectively, due to both the lack of atomic cooling
haloes of primordial composition and a lack of atomic cooling haloes
that rapidly grow.  The rare-peak region has a volume of 133.6
cMpc$^3$ (equivalent to a sphere of approximately 2.25 $h^{-1}$ cMpc
in radius) and an average density of 1.65 times the cosmic mean.  At
$z=15$, regions of that size and overdensity are expected to be found
in approximately $10^{-4} - 10^{-3}$ of the volume of the universe,
implying that a more realistic estimate of the global number density
of DCBH candidates forming through this mechanism is in the range of
$10^{-7} - 10^{-6}$ haloes/cMpc$^3$, consistent with our estimate in
the normal region.  This estimate is nonetheless approximately
100--1000 times greater than the observed $z \sim 6$ quasar number
density\cite{Onoue_2017}.  \change{Our results suggest that faint
  quasars at these high redshifts should be strongly clustered and
  associated with galaxy overdensities\cite{Ota18}, providing an
  observational test of our proposed seeding mechanism.}  Furthermore,
we find that DCBHs forming in this scenario have the same number
density as the observed number density\cite{Shankar_2009,
  Terrazas_2016} of present-day SMBHs above $10^8\Ms$.  This abundance
matching intriguingly suggests that such central black holes in most
elliptical galaxies have a common origin beginning their lives as
SMSs.

\renewcommand{\figurename}{{\bf Extended Data Table}}
\setcounter{figure}{0} 

\begin{table*}
  \caption{Properties of halo candidates hosting supermassive star formation}
  \label{tab:haloes}
  \centering
   \begin{tabular*}{0.99\textwidth}{@{\extracolsep{\fill}}c c c c c c}
     \hline
     log$_{10}$(M$_{\rm halo}$) & log$_{10}$(mean Growth rate)
     & J$_{\rm LW}$/J$_{21}$ & D$_{\rm gal}$ & T$_{\rm c}$
     & Gas infall rate\\

     [M$_\odot$] & [M$_\odot$ per unit redshift] & & [kpc] & [K] 
     & [M$_\odot$/yr]\\
     \hline
     7.84$^*$ & 7.78 & 2.71 & 12.7 & 2250 & 0.275\\  
     7.76$^\dag$ & 7.53 & 3.23 & 11.8 & 4390 & 0.171\\ 
     7.76 & 7.76 & 1.91 & 14.7 & 4220 & 0.286\\ 
     7.75 & 7.65 & 0.583 & 35.5 & 1730 & 0.290\\ 
     7.75 & 7.39 & 0.958 & 19.7 & 7570 & 0.0294\\ 
     7.74 & 7.88 & 1.49 & 25.0 & 8670 & 0.0574\\ 
     7.73 & 7.79 & 0.894 & 29.9 & 1760 & 0.396\\ 
     7.70 & 7.22 & 2.62 & 18.3 & 6520 & 0.0292\\ 
     7.67 & 7.90 & 0.16 & 124 & 1080 & 1.05\\ 
     7.64 & 7.78 & 2.14 & 6.20 & 7890 & 0.0356\\ 
     \hline
   \end{tabular*}
   \parbox[t]{0.99\textwidth}{\textit{Table notes:} 
     $^*$Most massive halo (MMH). $^\dag$Halo exposed to the highest
     Lyman-Werner radiation flux (LWH).  The growth rate is averaged over the last
     20 million years of the simulation.  The values of J$_{\rm LW}$,
     Lyman-Werner intensity in units of J$_{21}$, and T$_{\rm c}$, the
     gas temperature, are given at the densest point. D$_{\rm gal}$ is
     the distance to the nearest galaxy with at least $10^6~\Ms$ of
     stars.  The gas infall rate is the mass-averaged value within 100
     parsecs. All data are given at redshift 15 from the original
     Renaissance Simulation of the rare-peak region.
   }
\end{table*}

\textbf{The fraction of rapidly growing atomic cooling haloes.}  We
compute an additional check on the fraction of rapidly growing haloes
with semi-analytical galaxy formation code {\sc
  Galacticus}\cite{Benson_2012}.  Most of the rare-peak region
collapses into a halo with a total mass $8.5 \times 10^{11}\Ms$ at $z
= 5$.  We calculate 50 merger trees of equivalent $z=5$ haloes that
follow progenitor haloes as small as $3 \times 10^5\Ms$.  Each merger
tree has approximately 40,000 progenitor haloes at $z=15$, totaling 2
million progenitors in all of the calculated merger trees at this
epoch.  We find that $3 \times 10^{-4}$ of haloes around the atomic
cooling threshold, $M_{\rm halo} = (3-6) \times 10^7\Ms$, grow more
than a factor of six between redshifts 16 and 15 ($\sim$20~Myr).  By
using the halo descendant at a much later time, we can incorporate the
effects of being in an overdense environment that leads to higher halo
mass growth rates.  This leads to a more accurate estimate when
compared to randomly sampling atomic cooling haloes at $z=15$ whose
merger trees would be unlikely to be representative of our target
haloes.

\textbf{The metal-free haloes found in the Renaissance Simulations.}
We searched and have reported on the final output at $z=15$ of the
rare-peak region in the Renaissance simulation that satisfied the
following criteria: (1) an atomic cooling halo ($M_{\rm halo} \ge 4.9
\times 10^7~\Ms$ at $z=15$), (2) only contains high-resolution dark
matter particles, (3) does not support metal or dust radiative cooling
($Z < 10^{-6}~\textrm{Z}_\odot$), and (4) no prior star formation.  We
consider the last criterion because some Pop III stars directly
collapse into a stellar-mass black hole, producing no
metals\cite{Heger_2003, Chatzopoulos_2012}; however because we
randomly sample from an initial mass function, it is just as likely
that the same Pop III star particle in question could have chemically
enriched its host halo.  Out of 670 atomic cooling haloes, ten haloes
fit these conditions with their details shown in Extended Data Table
\ref{tab:haloes}.  We also searched the ``normal'' and ``void'' regions
and found no haloes matching the same criteria at $z=15$.  This
suggests that SMS formation through this channel is much more likely
in overdense regions of the early universe.  This crowded
high-redshift environment lends itself to high LW radiation field from
more nearby galaxies and haloes with very high growth rates, existing
in a medium with a higher average mass density.  This null detection
agrees with our statistical expectations discussed earlier.

We have focused on the most massive and most irradiated haloes in the
main article, but from the other eight haloes, we can gain insight on
their variations.  Eight of ten haloes experienced growth rate greater
than $10^{7.5}~\Ms$ per unit redshift, showing that they grow from the
minimum mass $M_{\rm crit}$ for molecular hydrogen cooling to one
supporting atomic cooling in less than a dynamical time
($\sim$20~Myr).  With the exception of one halo, they have a galaxy
with a stellar mass over $10^6~\Ms$ within 40 proper kpc.  Stars in
nearby galaxies create a boosted LW radiation field above $0.5
J_{21}$.  Inspecting the central temperatures, we see that there are
six cool core ($T < 5000$~K) haloes that have efficient H$_2$ cooling,
whereas the other four haloes still have warm cores.  For various
reasons, such as the LW radiation field and the degree of turbulence
in the core\cite{Wise_2007b}, the warm core haloes have their cooling
suppressed at the simulation's final redshift.  However, this state is
temporary, and the system will cool within a timescale defined by the
ratio of the H$_2$ fraction and its formation rate as the haloes
continue to grow\cite{Wise_2007b}.  The higher core pressure limits
the gravitational collapse, resulting in lower infall rates
($<0.1$~\Msyr{}) when compared to the cool core haloes.  The infall
rates in our SMS candidate list do not follow the usual thermal
accretion rate ($c_{\rm s}^3/G$) because they systems have lost their
thermal support in a short timescale relative to the free-fall time.
Thus, they are experiencing their initial catastrophic collapse.
Although SMS models have shown that a critical infall rate of
0.04~\Msyr{} is required, we include these warm core haloes in our
candidate list because molecular cooling and thus a further and more
rapid collapse is inevitable\cite{Visbal_2014}.  The halo with the
highest infall rate ($\sim 1$~\Msyr{}) is an interesting case because
it exists the farthest (124 proper kpc) from a galaxy and thus has the
lowest $J_{\rm LW}$ value in this sample.  This halo also experiences
the most rapid growth rate, which we attribute to it not collapsing
earlier before it reaches the atomic cooling limit.  We also note here
that the accretion rates given are for a single snapshot, and the
average accretion rate will vary as such haloes undergo quiescent and
intensive periods of cosmological accretion.

\textbf{Data Availability.}
The numerical experiments presented in this work were run with a
hybrid OpenMP+MPI fork of the {\sc Enzo} code available from
\url{https://bitbucket.org/jwise77/openmp}, using the changeset
bcb436949d16.  The data are publicly available from the Renaissance
Simulation Laboratory at \url{http://girder.rensimlab.xyz}.

\end{methods}


\newpage








\end{document}